\documentclass[12pt]{article}
\pdfoutput=1
\usepackage{amsmath,amssymb} 
\usepackage{slashed}
\usepackage{color}
\usepackage{cite}
\usepackage{graphicx}
\usepackage{hyperref}
\usepackage{subdepth}

\usepackage[margin=3cm]{geometry}
\numberwithin{equation}{section}

\usepackage[small,bf,font=small,labelsep=period]{caption}

\newcommand{\nn}{\nonumber\\}

\def\tilde{\widetilde}
\def\CO{\mathcal O}

\def\sfU{\mathsf{U}}

\def\sfM{\mathsf{M}}

\def\ImO{\mathrm{Im}\mathcal{O}}

\def\dt{\tilde{d}}
\def\ddt{{\tilde{d}\,}^2}
\newcommand{\im}[1]{\mathrm{Im} {#1}}

\DeclareRobustCommand{\bfrac}[2]{%
	\mathchoice{\frac{\raisebox{-0.4ex}{$#1$}}{\raisebox{-0.1ex}{$#2$}}}%
	{\frac{\raisebox{-0.3ex}{$\scriptstyle#1$}}{\raisebox{-0.4ex}{$\scriptstyle#2$}}}%
	{\frac{#1}{#2}}%
	{\frac{#1}{#2}}%
	}

\catcode`@=12

\begin{document}
\begin{titlepage}

\begin{center}
\parbox[t]{\textwidth}{\centering \LARGE	\bf
	\vspace{1em}
	Symmetry breaking in holographic \\
	theories with Lifshitz scaling
	}

\vspace{2em}

{\large 
		Riccardo Argurio$^{1}$, Jelle Hartong$^{2}$,\\ 
		\vspace{1ex}Andrea Marzolla$^{1}$, and Daniel Naegels$^{1}$}

\vspace{2em}
\parbox[t]{0.8\textwidth}{\itshape \linespread{1} \footnotesize \centering
{$^1$ Physique Th\'eorique et Math\'ematique and International Solvay Institutes, Universit\'e Libre de Bruxelles, C.P. 231, 1050 Brussels, Belgium.}

\vspace{1ex}
{$^2$ Institute for Theoretical Physics and Delta Institute for Theoretical Physics, University of Amsterdam, Science Park 904, 1098 XH Amsterdam, \\The Netherlands.}

}

\end{center}

\vspace{3em}

\noindent
\begin{center}
\bf \small Abstract
\end{center}
\noindent
We study holographically Lifshitz-scaling theories with broken symmetries. In order to do this, we set up a bulk action with a complex scalar and a massless vector on a background which consists in a Lifshitz metric and a massive vector. We first study separately the complex scalar and the massless vector, finding a similar pattern in the two-point functions that we can compute analytically. 
By coupling the probe complex scalar to the background massive vector we can construct probe actions that are more general than the usual Klein--Gordon action. Some of these actions have Galilean boost symmetry. Finally, in the presence of a symmetry breaking scalar profile in the bulk, we reproduce the expected Ward identities of a Lifshitz-scaling theory with a broken global continuous symmetry. In the spontaneous case, the latter imply the presence of a gapless mode, the Goldstone boson, which will have dispersion relations dictated by the Lifshitz scaling.

\end{titlepage}

\tableofcontents

\section{Introduction}

Our goal is to discuss symmetry breaking and Goldstone bosons in theories that enjoy non-relativistic (Lifshitz) scale invariance. Low-energy effective theories for such Goldstone bosons were considered for instance in \cite{Griffin:2013dfa,Griffin:2014bta,Griffin:2015hxa,Horava:2016vkl}. If one is interested in seeing how such Goldstone bosons arise in strongly coupled theories, the natural setting is to turn to large $N$ field theories and their holographic description. Holography for theories with Lifshitz scale symmetry has been initiated in \cite{Kachru:2008yh,Taylor:2008tg} (see \cite{Taylor:2015glc} for a recent review). Some other relevant references are \cite{Baggio:2011cp,Ross:2011gu,Andrade:2012xy,Korovin:2013bua,Keeler:2013msa,Christensen:2013rfa,Keeler:2014lia,Chemissany:2014xsa,Hartong:2014oma,Hartong:2015wxa,Sybesma:2015oha,Keeler:2015afa,Keranen:2016ija}.

In this paper, we will be concerned with the breaking of a global symmetry which commutes with the spacetime symmetries of the theory. This entails the presence of a conserved current (i.e. a charge density and a spatial current related by a conservation equation), independent from the energy-momentum tensor. Holographically, this means that we will have to deal with a bulk massless vector in a fixed Lifshitz background. We will consider Lifshitz backgrounds that are a solution of Einstein gravity coupled to a massive bulk vector field which acquires a non-trivial Lifshitz invariant profile. However, we will not consider the dynamics, neither of the metric nor of this massive vector field, and instead treat them as fixed background quantities.

We will consider the dynamics of a massless vector and a charged scalar in the bulk. When the latter acquires a profile, it corresponds to the dual scalar operator breaking the continuous symmetry by a vacuum expectation value (or an explicit breaking term). The probe complex scalar and the massless vector field couple also to the background quantities: not only the Lifshitz metric but also the massive vector field. By including all possible couplings to the latter, we will be able to consider bulk actions for the scalar and the massless vector that are more general than the standard Klein-Gordon and Maxwell actions.

Since in order to discuss symmetry breaking, we need to consider the coupled system of a scalar and a vector, we will need first to review how Lifshitz holography works for a scalar and a massless vector separately. Indeed, having less constraints coming from Lifshitz scale symmetry as opposed to conformal symmetry, the field theory outcome from holography is not completely fixed by kinematics, and it moreover depends on choices that one makes in the bulk theory. 

We thus start in Section~\ref{scalarLif} by recalling known facts about holography for a scalar on a Lifshitz bulk background. We extract analytical expressions for the two-point functions when it is possible, and comment on their physical meaning, also exploring a few generalizations. Then in Section~\ref{vectorLif} we turn to massless vectors, which received surprisingly little attention in Lifshitz holography. We comment on the characteristics of the dual conserved current. In Section~\ref{GoldstoneLif} we finally address the coupled system, holographically deriving the Ward identities of non-relativistic symmetry breaking. In the Appendix, we provide some comments on low-energy field theories of Lifshitz Goldstone bosons, and the associated Ward identities.

\section{Lifshitz holography for a scalar, revisited}\label{scalarLif}

Lifshitz symmetry is the invariance under an anisotropic scaling of time and spatial coordinates,
\begin{equation}	\label{lifscaling}
\begin{aligned}
& t 				&\longrightarrow&\quad \lambda^z\,t \ ,\nn
& x_i 	&\longrightarrow&\quad \lambda\,x_i \ , \nonumber
\end{aligned}
\end{equation}
where $z$ is called the dynamical critical exponent. Whenever $z\neq 1$, such scaling transformations are not compatible with Lorentz symmetry, and they constitute non-relativistic scale transformations. We will always assume that $z\ge 1$. If we add to this scale symmetry, time and space translations, as well as spatial rotations, we obtain what is called the Lifshitz symmetry group. 

A $d+1$-dimensional Lifshitz invariant metric with $SO(d-1)$ rotational symmetries is unique and takes the following form
\begin{equation}	\label{lifmetric}
ds^2 = \bfrac{dr^2}{r^2} -\bfrac{dt^2}{r^{2z}} +\bfrac{\sum_idx_i^2}{r^2}\,.
\end{equation}
It can be shown that such a metric for $z\neq 1$ is a solution of a suitably chosen theory of Einstein gravity with a negative cosmological constant and a massive vector field \cite{Taylor:2008tg}, which we will denote as $B$ and which is given by
\begin{equation}\label{Lifvector}
B=\beta\frac{dt}{r^z}\ , \quad	\text{with }\ \beta\equiv\sqrt{\frac{2(z-1)}{z}} \ .
\end{equation}
For $z=1$ we recover the standard AdS metric in Poincar\'e coordinates that has relativistic scale invariance. The spatial coordinate~$r$ will be our holographic coordinate that is zero at the boundary and that tends to infinity in the bulk.

In the following we will consider the metric~\eqref{lifmetric} and the massive vector~\eqref{Lifvector} as fixed, non-dynamical background quantities, and we will study the holographic renormalization and boundary two-point functions of operators dual to fields on a fixed Lifshitz bulk geometry. We start reviewing the properties of a scalar field on a Lifshitz background.

\subsection{Two-point function for Lifshitz Klein-Gordon scalar}	\label{2pscalarLif}

The basic setup of a free scalar field in Lifshitz spacetime has of course been considered previously in the literature~\cite{Kachru:2008yh,Taylor:2008tg,Taylor:2015glc}. This simple case in fact already displays some peculiar features with respect to the AdS version, namely the appearance of imaginary poles in the two-point function. These diffusive poles at zero temperature are a characteristic feature in holographic Lifshitz theories, and they will appear also in the vector sector as we will show in Section~\ref{vectorLif2p}. In Section~\ref{salarLifHW} we will explore the possible relation between these diffusive poles and the presence of a tidal singularity in the center of the Lifshitz spacetime, by using ``hard wall'' boundary conditions in the bulk in order to cut out the singularity.

Let us begin by considering the $(d\!+\!1)$-dimensional bulk Klein-Gordon action for a complex scalar field
\begin{equation}	\label{Sscalar}
	S=\int\!d^{d+1}x\ \sqrt{-g\,}\,\Big( -g^{\mu\nu}\partial_{\mu}\phi^*\partial_{\nu}\phi-m^2\,\phi^*\phi \Big) \ ,
\end{equation}
where $g$ is the Lifshitz metric, as defined by~\eqref{lifmetric}. We allow for arbitrary values of the bulk mass~$m$, but, for simplicity, we do not consider higher order potential terms.

From the variation of this action, we obtain the equation of motion
\begin{equation}	\label{EOMscalar}
r\partial_r(r\partial_r\phi)-\left(d+z-1\right) r\partial_r\phi -r^{2z}\partial_{t}^{2}\phi+r^{2}\partial_{i}^2\phi-m^{2}\phi=0\ ,
\end{equation}
where $\partial_i^2$ implies a sum over the $d\!-\!1$ values of~$i$. This equation has two independent solutions, with the following near boundary asymptotics ($r=0$):
\begin{equation}	\label{asymphi}
\phi = r^{\frac{\tilde{d}}{2}-\nu}\,\big(\phi_0+\ldots\big) +r^{\frac{\tilde{d}}{2}+\nu}\,\big(\tilde{\phi}_0+\ldots\big)\ , \qquad
	\text{with }\ \left\{	
						\begin{array}{l}
\vphantom{\Big|}		\dt=d+z-1 \ , \\
\displaystyle			\nu=\sqrt{\bfrac{\,\ddt}{4}+m^2}\ .
						\end{array}\right.\!\!\!\!\!
\end{equation}
When $\nu\!=\!0$ we have a Lifshitz version of the BF (Breitenlohner-Freedman) bound~\cite{Breitenlohner:1982jf} on the bulk mass of the scalar, with an `effective' dimension~$\dt$, obtained by adding the dynamical critical exponent $z$ to the number of spatial boundary dimensions $d-1$ (that is, counting the time dimension $z$~times). From now on, in order to avoid technical (and not crucial) difficulties related to the renormalization procedure, we will take $0\!<\!\nu\!<\!1$.

Using the equations of motion~\eqref{EOMscalar} we can put the action~\eqref{Sscalar} on-shell, and reduce it to a boundary term,
\begin{align}
S_\mathrm{on\text{-}shell} &=
	\frac12\int\!d^{d+1}x\ \partial_r\Big[ -r^{-\dt}\big( \phi^*\,r\partial_r\phi +\phi\,r\partial_r\phi^* \big)\Big] 	\label{SscalarOS}\\
&=	
	\int_{r=\epsilon}\!d^dx\ \bigg[ \bigg(\bfrac{\dt}{2}-\nu\bigg)r^{-2\nu}\,\phi_0^*\phi_0 +\bfrac{\dt}{2}\Big(\phi_0\tilde{\phi}_0^*+\phi_0^*\tilde{\phi}_0\Big) \bigg] \ , \nonumber
\end{align}
where in the second line we have used the asymptotic expansion~\eqref{asymphi}, and, since we have a term diverging at the boundary~$r\!=\!0$, we have kept a small~$\epsilon$ as a regulator. We can renormalize such a divergence through a standard mass counterterm
\begin{equation}\label{Sctmassscalar}
S_{ct} = 
	\bigg(\bfrac{\dt}{2}-\nu\bigg)\int_{r=\epsilon}\!d^{d}x\;\sqrt{-\hat{g}\,}\:\phi^*\phi\ ,
\end{equation}
where $\hat{g}$ is the induced metric on the boundary, thus obtaining
\begin{equation}	\label{SrenScalar}
S_{ren}=S_\mathrm{on\text{-}shell}-S_{ct}= \nu\! \int\!d^{d}x\; \big(\phi_0\tilde{\phi}_0^*+\phi_0^*\tilde{\phi}_0\big)\ .
\end{equation}
A full solution to the equation of motion~\eqref{EOMscalar} would relate the subleading~$\tilde{\phi}_0$ to the leading~$\phi_0$ through a non-local function $f_\phi$, \emph{i.e.}~$\tilde{\phi}_0\!=\!f_\phi(\partial_t,\partial_i^2)\,\phi_0$. We assume this non-local function $f_\phi$ to be real. Then we can write the formula for the two-point correlator of the dual boundary operator~$\mathcal{O}_\phi$:
\begin{equation}	\label{2Pformula}
\big\langle\mathcal{O}_{\phi}(x)\,\mathcal{O}^*_{\phi}(x')\big\rangle= 
	-i\:\bfrac{\delta^2\;S_{ren}\phantom{\phi(0)\phi}}{\delta\phi_0(x)\delta\phi_0^*(x')} = 
		-i\:2\nu\;f_\phi(\partial_t,\partial_i^2)\;\delta(x-x')\ .
\end{equation}

The equation of motion~\eqref{EOMscalar} can be analytically solved for~$z\!=\!2$. Writing
\begin{equation}
\phi(r,t,x)=e^{-i\omega t+i\vec k\cdot\vec x}\phi(r,\omega,k)
\end{equation} 
we find for the Fourier modes
\begin{equation}	\label{EOMscalarFt}
r\partial_r(r\partial_r\phi)-\left(d+1\right)r\partial_r\phi-\big(m^{2}+k^2r^2-\omega^2r^4\big)\phi=0\ .
\end{equation}
The solution is given in terms of confluent hypergeometric functions, of the first kind,~$\sfM$, and of the second kind,~$\sfU$:
\begin{align}
\phi(r,\omega,k^2)=e^{\frac{i}{2}\omega\,r^{2}} r^{\bfrac{d+1}{2}+\nu}\, \Big( 
	C_{1}\:\sfU\big[a,1+\nu;-i\omega\,r^2\big] +
		C_{2}\:\sfM\big[a,1+\nu;-i\omega\,r^2\big]\Big)\ , \label{solScalar}\\
\text{with }\ 
	a=\frac{1}{2}\left(1+\nu+\frac{ik^2}{2\omega}\right)\ .	\label{adef}
\end{align}
We now have to impose a boundary condition in the bulk, in order to fix one of the two integration constants. We choose to select in-going waves in the extreme bulk ($r\to \infty$),\footnote{
	We are adopting the usual prescription of~\cite{Son:2002sd}, which is actually equivalent to working in Euclidean signature. A more rigorous derivation of the analytic structure of the time-ordered Minkowski correlator should be recovered by accurately performing real-time holography~\cite{Skenderis:2008dh,Skenderis:2008dg}.}
which is given by setting~$C_2=0$. Indeed the function $\sfM\big[a,1+\nu;-i\omega\,r^2\big]$ for large argument goes as~$e^{-i\omega\,r^{2}}$, thus introducing an out-going wave. Note that the Fourier mode in~\eqref{solScalar}, when $C_2$ is set to zero, decays in the deep bulk for~$\im{\omega}\!>\!0$.  Wick-rotating to Euclidean signature, this corresponds to requiring regularity in the deep bulk.

From the expansion of $\sfU[a,b;x]$ around~$x\!=\!0$,
\begin{equation}	\label{Usmallx}
\sfU[a,b;x]= \bfrac{\Gamma[1-b]}{\Gamma[a-b+1]}+\dots +x^{1-b}\,\bfrac{\Gamma[b-1]}{\Gamma[a]}+\dots \ ,
\end{equation}
we can read off the coefficient of the leading term, going as $r^{\frac{d+1}{2}-\nu}$, which is~$\phi_0$, and the coefficient of the subleading term, going as $r^{\frac{d+1}{2}+\nu}$, which is~$\tilde{\phi}_0$. Then from the formula~\eqref{2Pformula} we obtain the two-point function of the dual scalar operator with Lifshitz scaling~$z\!=\!2$:
\begin{align}
\big\langle\mathcal{O}_{\phi}(k)\,\mathcal{O}^*_{\phi}(-k)\big\rangle &=
	-i\:2\nu\;f_\phi(\omega,k_i^2) = -i\:2\nu\;\frac{\tilde{\phi}_0}{\phi_0} \nn
&=
	2i(-i\omega)^\nu\; \frac{\Gamma[1-\nu]}{\Gamma[\nu]}\, \frac{\Gamma\Big[\frac{1}{2}\Big(1+\nu+\frac{ik^2}{2\omega}\Big)\Big]}{\Gamma\Big[\frac{1}{2}\Big(1-\nu+\frac{ik^2}{2\omega}\Big)\Big]}\ .	
\label{scalarLifcorr}
\end{align}

We first remark the branch-cut starting at $\omega=0$. Then, the Euler gamma function~$\Gamma$ has poles at negative integer values of its argument and $1/\Gamma$ is an entire function, so we find that the poles of the correlator are given by
\begin{equation}	\label{immpoles}
\omega=-\bfrac{ik^{2}}{2(2n +1 +\nu)}\ , \quad \text{ with }n\in \mathbb{N}\,.
\end{equation}
We thus see that these poles all lie on the negative imaginary axis.\footnote{We notice that the imaginary poles are symmetric under~$\omega\!\to\!-\omega^*$, that is the equivalent condition in the complex~$\omega$ plane to time-reversal invariance, which is indeed a symmetry of Lifshitz Klein-Gordon equation~\eqref{EOMscalarFt}. The branch-cut in~\eqref{scalarLifcorr} may also be taken to be symmetric under~$\omega\!\to\!-\omega^*$, i.e.~lying along the upper imaginary axis.}
In fact the locations of the poles accumulate as one approaches the origin along the negative imaginary axis. One may expect this behavior to be a consequence of the singularity at $r\!=\!\infty$ where for $z\!>\!1$ Lifshitz spacetimes display diverging tidal forces~\cite{Kachru:2008yh}. In other words Lifshitz spacetimes with $z\!>\!1$ are, contrary to AdS, geodesically incomplete. In the next section we make a little detour in order to check if cutting out this singularity, by using ``hard wall'' boundary conditions, gets rid of the imaginary poles. For other considerations on the appearance of imaginary poles in Lifshitz holography, together with a field theoretic perspective on the issue, see \cite{Keeler:2013msa,Keeler:2014lia,Keeler:2015afa}.

\subsection{The hard wall solution for Lifshitz Klein-Gordon scalar}	\label{salarLifHW}

We want to see how the two-point function, and its poles, change if we break the Lifshitz scale invariance setting a hard wall in the IR, i.e. deep in the bulk.

We consider exactly the same bulk action~\eqref{Sscalar} as in the previous section, so we have the same solution of the equation of motion~\eqref{solScalar} for~$z\!=\!2$. But now we impose a different IR regularity condition, namely that the field vanishes\footnote{This IR boundary condition might seem closer in spirit to the ones used after a Euclidean rotation. However as we have noted, IR vanishing and in-falling boundary conditions coincide when going to the upper half $\omega$ plane. See also \cite{Andrade:2012xy} for a similar situation.} in the bulk on a surface at $r\!=\!\mu^{-1}$. This yields
\begin{equation}
\phi^{HW}(\mu^{-1},\omega,k^2) \equiv 0 \quad\Leftrightarrow\quad 
	C_2 = -C_1\; \bfrac{\sfU\Big[a,1+\nu;-\frac{i\omega}{\mu^{2}}\Big]}{\sfM\Big[a,1+\nu;-\frac{i\omega}{\mu^{2}}\Big]}\ ,
\end{equation}
with the expression for $a$ already given in~\eqref{adef}, so that the hard wall solution reads
\begin{align}
&	\phi^{HW}(r,\omega,k^2) = \\
&	\quad =
	C_1 e^{\frac{i}{2}\omega\,r^{2}} r^{\frac{d+1}{2}+\nu} \bigg( 
\sfU\big[a,1+\nu;-i\omega\,r^2\big] -\bfrac{\sfU\big[a,1+\nu;-\mu^{-2}i\omega\big]}{\sfM\big[a,1+\nu;-\mu^{-2}i\omega\big]}\, \sfM\big[a,1+\nu;-i\omega\,r^2\big] \!\bigg)\ , \nonumber
\end{align}
and $\phi^{HW}\!\equiv\!0$ for $r\!>\!\mu^{-1}$.

In order to determine the two-point correlator, we consider the terms $\propto\!r^{\frac{d+1}{2}-\nu}$ and $\propto\!r^{\frac{d+1}{2}+\nu}$ in the series expansion of the solution~$\phi^{HW}$ around $r\!=\!0$. Using the near boundary expansion~\eqref{Usmallx} for the confluent hypergeometric function of the second kind $\sfU\big[a,b;x\big]$, also known as the Tricomi function, and recalling that~$\sfM[a,b;x]\!=\!1 +O(x)$ for small~$x$, we find 
\begin{equation}
\begin{aligned}
\phi^{HW}_0 & = C_1\, \left(-i\omega\right)^{-\nu} \frac{\Gamma[\nu]}{\Gamma[a]}\ ,	\\
	\tilde{\phi}^{HW}_0 & = C_1\, \bigg( \frac{\Gamma[-\nu]}{\Gamma[a-\nu]} -\bfrac{\sfU\big[a,1+\nu;-\mu^{-2}i\omega\big]}{\sfM\big[a,1+\nu;-\mu^{-2}i\omega\big]} \bigg)\ .
\end{aligned}
\end{equation}
The correlator of the dual scalar operator, following formula~\eqref{2Pformula}, is then given by
\begin{align}
\langle O_\phi O_\phi^*\rangle_{HW} = -i\,2\nu\;\frac{\tilde{\phi}^{HW}_0}{\phi^{HW}_0} &=
	 -i\,2\nu\:\left(-i\omega\right)^{\nu}\: \frac{\Gamma[a]}{\Gamma[\nu]}\,	\bigg( \frac{\Gamma[-\nu]}{\Gamma[a-\nu]}  -\bfrac{\sfU\big[a,1+\nu;\,-\mu^{-2}i\omega\big]}{\sfM\big[a,1+\nu;\,-\mu^{-2}i\omega\big]} \bigg) \nn
&=
	i\:2\nu\,\mu^{2\nu}\; \bfrac{\sfM\big[a-\nu,1-\nu;\,-\mu^{-2}i\omega\big]}{\sfM\big[a,1+\nu;\,-\mu^{-2}i\omega\big]}\ , 	\label{HWcorr}
\end{align}
where in the second line we have used the expression of the Tricomi function as a combination of confluent hypergeometric functions, that is
\begin{equation*}
\sfU[a,b;x]= \bfrac{\Gamma[1-b]}{\Gamma[a+1-b]}\;\sfM[a,b;x] +x^{1-b}\;\bfrac{\Gamma[b-1]}{\Gamma[a]}\;\sfM[a+1-b,2-b;x]\ .
\end{equation*}
Notice that this simplification precisely removes the branch cut on the real axis which was present in the pure Lifshitz correlator~\eqref{scalarLifcorr}. 

\begin{figure}[bt]
	\centering
	\begin{minipage}{0.9\textwidth}
		\flushright
		\small$i\langle O_\phi O_\phi^*\rangle_{HW}\hspace{52mm}$
		\vskip0.5ex
		\includegraphics[width=0.91\textwidth]{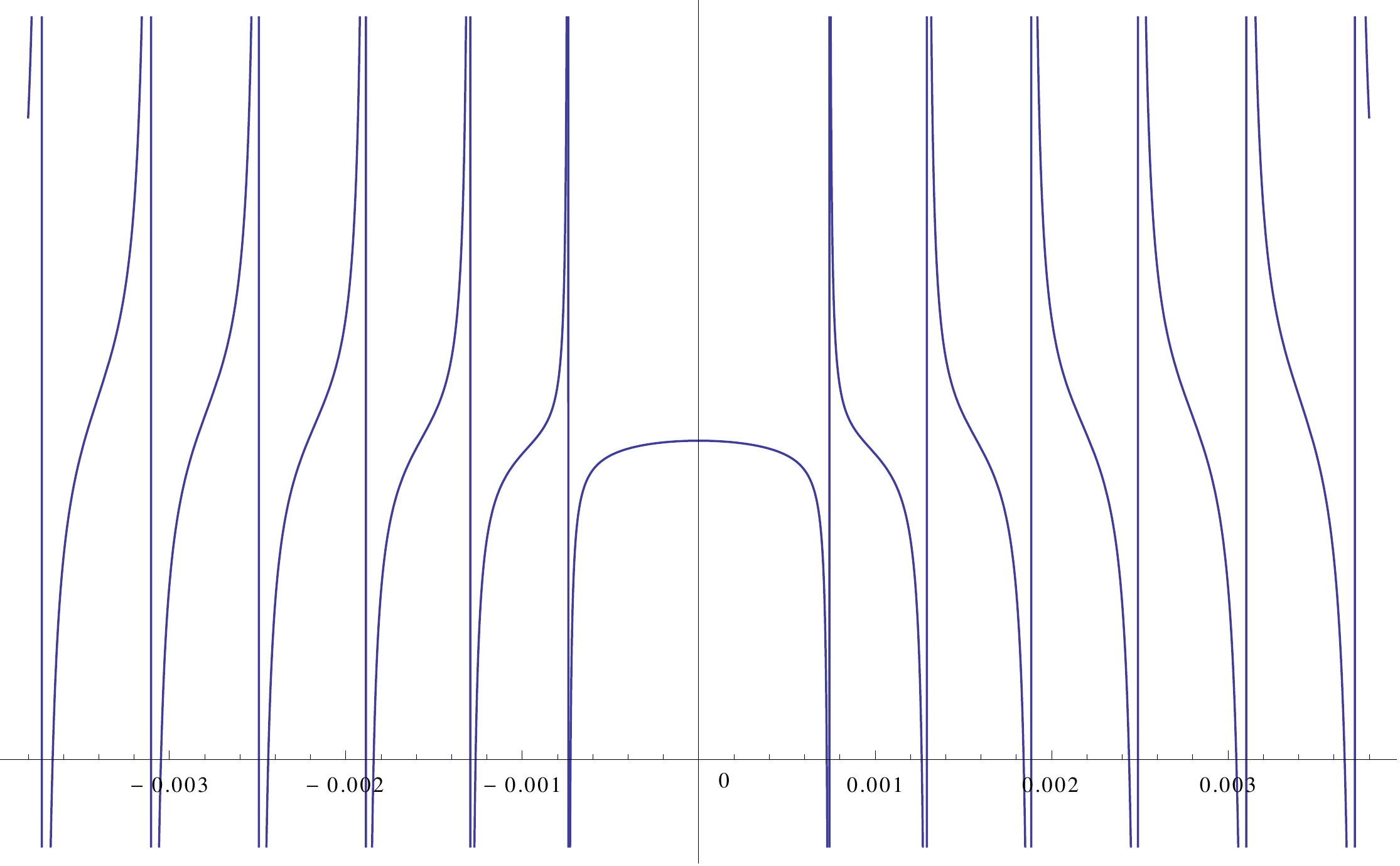}
	\end{minipage}
	\begin{minipage}{0.09\textwidth}
		\flushleft\vspace{65mm}
		\small$\omega$
	\end{minipage}
	\begin{minipage}[t]{0.9\textwidth}
		\centering
		\caption{Plot of the hard-wall correlator for \emph{real} values of~$\omega$, at fixed~$\mu=0.01$, $\nu=\tfrac{1}{4}$, $k^2=0.001$.}
		\label{LifHWrealpoles}
	\end{minipage}
\end{figure}
\begin{figure}[bt]
\centering
\begin{minipage}{0.03\textwidth}
\end{minipage}
\begin{minipage}{0.45\textwidth}
\centering
\scriptsize$\quad i\langle O_\phi O_\phi^*\rangle_{HW}$
\vskip1.6ex
\includegraphics[width=\textwidth]{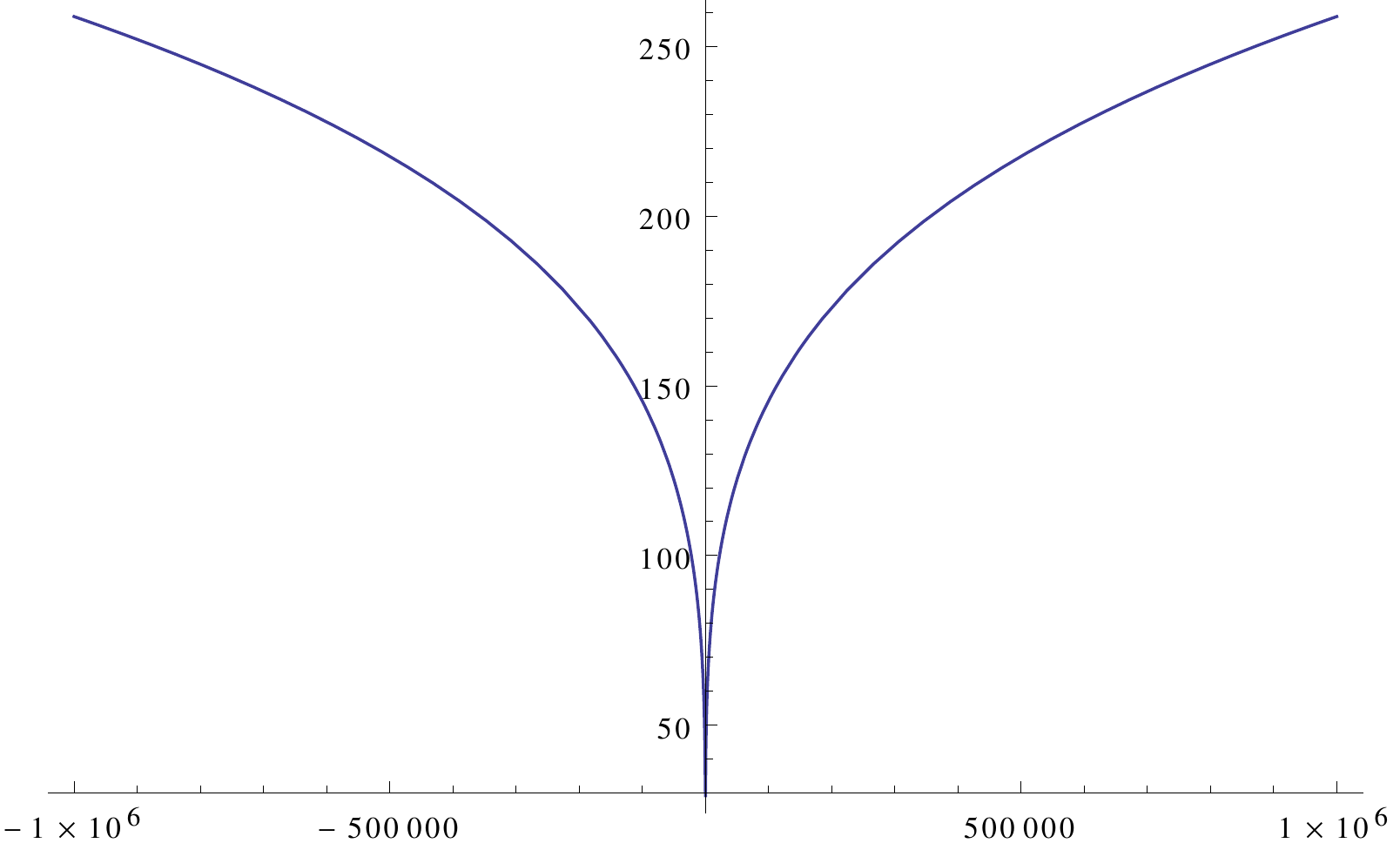}
\end{minipage}
\begin{minipage}{0.03\textwidth}
\flushleft\vspace{40mm}
\scriptsize$\!\!-i\omega$
\end{minipage}
\begin{minipage}{0.45\textwidth}
\centering
\scriptsize$\quad i\langle O_\phi O_\phi^*\rangle_{HW}$
\vskip0.3ex
\includegraphics[width=0.99\textwidth]{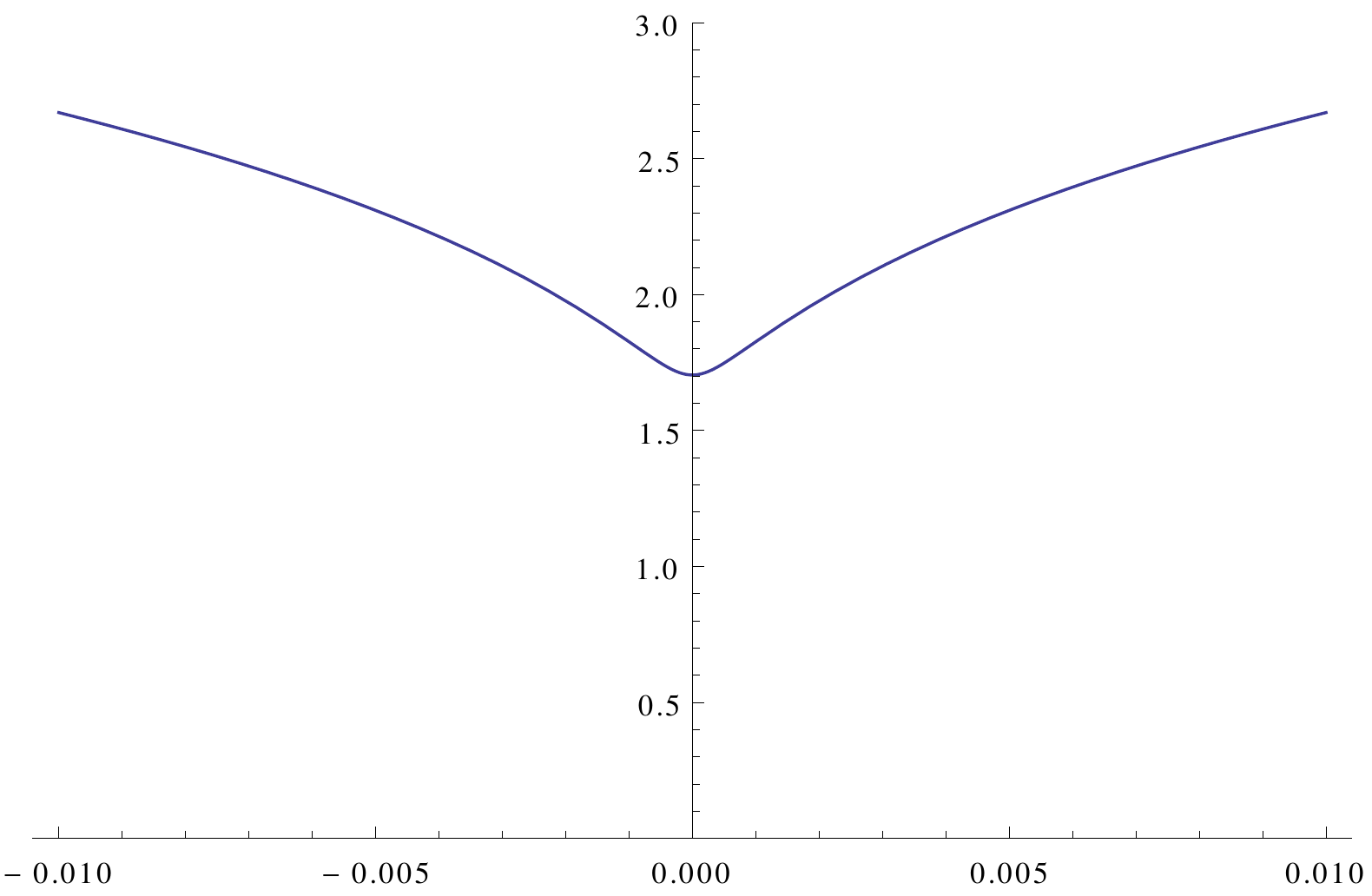}
\end{minipage}
\begin{minipage}{0.03\textwidth}
\flushleft\vspace{41mm}
\scriptsize$\!\!-i\omega$
\end{minipage}
\begin{minipage}[t]{0.9\textwidth}
\vskip1ex
\caption{Plot of the hard-wall correlator for purely \emph{imaginary} values of~$\omega$, at fixed~$\mu=0.01$, $\nu=\tfrac{1}{4}$, $k^2=0.001$. The second plot is a zoom close to the origin of the first plot.}
\label{LifHWpureImOmega}
\end{minipage}
\end{figure}

Note that $a$ depends on $\omega$ so that the frequency dependence of $\sfM[a,b;x]$ is through both $a$ and $x$. Since the confluent hypergeometric function has no poles, the poles of the hard-wall correlator are determined by the zeros of $\sfM$. Since zeros of analytic functions lie isolated we obtain a discrete spectrum as shown in the numerical plot in Fig.~\ref{LifHWrealpoles}, for the case of poles that lie on the real-$\omega$ axis. The zeros of $\sfM$ do not have a simple analytic expression and are hard to find even numerically. The numerical plot in Fig.~\ref{LifHWpureImOmega} shows that the correlator~\eqref{HWcorr} has no poles that lie exactly on the imaginary axis, as in the pure Lifshitz case~\eqref{immpoles}, but it should have poles that lie off the real and imaginary axes, in the complex plane, that in the $\mu\to0$~limit (no hard wall) align on the imaginary axis, at the locations of~\eqref{immpoles}. Thus we have a strong indication that the presence of complex poles in Lifshtz holography at zero temperature should not be attributed to the tidal singularity.

In addition, we remark that the limit of the correlator~\eqref{HWcorr} for $\mu\!\to\!0$ correctly gives the expression for the scalar correlator in the absence of the hard wall~\eqref{scalarLifcorr}, provided we approach the origin from the half-plane where $\im\,{\omega}\!>\!0$ (in agreement with the choice we made below \eqref{adef}). In fact, from the first line of~\eqref{HWcorr}, one can see that the correlator of the hard wall scalar is written in terms of the correlator of the pure Klein-Gordon scalar, plus a piece proportional to~$\sfU/\sfM$. In the limit $\mu\!\to\!0$, so large third argument, we have $\sfU[a,b;x]\propto x^a$, while $\sfM[a,b;x]\!\propto e^x$, with $x\!\equiv\!-\mu^{-2}i\omega$. So, for $\im\,{\omega}\!>\!0$, the $\sfU/\sfM$ piece is exponentially (and non-analytically) suppressed as we push the hard wall into the deep bulk, eventually recovering the pure Klein-Gordon expression in the strict $\mu\!=\!0$ limit.

\subsection{Two-point function for a general scalar} 
\label{moregeneralscalar}

We discuss here a modification of the Klein-Gordon scalar bulk action \eqref{Sscalar}, by allowing couplings to the background massive vector field \eqref{Lifvector}. This will allow for a time-reversal breaking kinetic term inspired by~\cite{Hartong:2015wxa}, which yields a more general model for a non-relativistic probe scalar, including a case with enhanced $z=2$ Schr\"odinger symmetry.

The background massive vector is given by equation \eqref{Lifvector}. Note that since $B_\mu B^\mu\!=\!-\beta^2$ we can define the metric orthogonal to $B_\mu$ as 
\begin{equation}	\label{orthometric}
\gamma_{\mu\nu}= g_{\mu\nu}+\beta^{-2}B_\mu B_\nu\ .
\end{equation}
We will denote by $\gamma^{\mu\nu}$ the above projector with its indices raised with $g^{\mu\nu}$.

Taking into account the fact that $B_\mu$, being massive, has no gauge invariance, the most general bulk action for a complex scalar is the following:
\begin{align}\label{Smoregenscal}
S=\int\!d^{d+1}x\; \sqrt{-g\,}\: \bigg[\,
	\bfrac1{c^2\beta^2}\,B^\mu\partial_\mu\phi^*B^\nu\partial_\nu\phi -\bfrac{ih}{2\beta}B^\mu\big(\phi^*\partial_\mu\phi-\phi\,\partial_\mu\phi^*\big)\: +\ & \nn
		-\,\gamma^{\mu\nu}\partial_\mu\phi^*\partial_\nu\phi -m^2 \phi^*\phi \, &\bigg] \ ,
\end{align}
where $c$ and $h$ are real numbers. Note that in order to have the second term, with the $h$ coefficient, we need a complex scalar for otherwise any term linear in $B^\mu \partial_\mu$ would be a total derivative. The contribution of the second kinetic term with coefficient $h$ to the kinetic energy is positive for positive frequency modes provided we take $h$ to be positive. We will therefore assume throughout that $h\ge 0$. We recover the Lifshitz Klein-Gordon case of the previous two subsections for $h\!=\!0$ and $c\!=\!1$, whereas the Schr\"odinger invariant case is recovered for $c\!=\!\infty$ as we will see later.

Since $B_\mu$ has no radial component, the terms involving $B_\mu$ in~\eqref{Smoregenscal} do not contribute to the on-shell boundary action, which is thus the same as in~\eqref{SscalarOS}. Thus, in our case, with $0<\nu<1$, the counterterm~\eqref{Sctmassscalar} removes the only divergence, so that the renormalized action is again given by~\eqref{SrenScalar}. Therefore, the two-point function is given by~\eqref{2Pformula}, where the unknown non-local function is now obtained by solving the equation of motion resulting from the variation of the generalized action~\eqref{Smoregenscal}, which is
\begin{equation}
r\partial_r(r\partial_r\phi) -\tilde{d}\,r\partial_r\phi +r^2\partial_i^2\phi -c^{-2}r^{2z}\partial_{t}^2\phi +ih\,r^z\partial_t\phi  -m^2\phi =0 \ .
\label{geneq}
\end{equation}
In order to find analytical results, once again we specialize to $z\!=\!2$. The above equation, after Fourier transforming, is identical to the Lifshitz Klein-Gordon equation~\eqref{EOMscalarFt}, with $\omega$ replaced by $\omega/c$ and $k^2$ replaced by $\tilde{k}^{\,2}\!=k^2\!-\!h\omega$:
\begin{equation}
r\partial_r(r\partial_r\phi) -\left(d+1\right)r\partial_r\phi +\bfrac{\omega^2}{c^2}\,r^4\phi -\tilde{k}^{\,2}r^2\phi -m^2\phi =0 \ .
\label{geneqz2}
\end{equation}
For $c\rightarrow\infty$ this equation is identical to that of a complex scalar on a Schr\"odinger spacetime with $z=2$ \cite{Balasubramanian:2008dm} and hence possesses $z=2$ Schr\"odinger invariance (see also \cite{Hartong:2015wxa}).\footnote{We note however that this is no longer true for other values of $z$, i.e. a complex scalar field on a Schr\"odinger spacetime with $z\neq 2$ satisfies a different equation than we would find here for $z\neq 2$ and $c\rightarrow\infty$.} The solution, again picking the in-falling part  as~$r\!\to\!\infty$, is given by
\begin{align}
\qquad \phi(r,\omega,k^2) = C\: e^{\frac{i\omega}{2c}\,r^2} r^{\frac{d+1}{2}+\nu}\; \sfU\Big[\,\tilde{a},1+\nu;-\frac{i\omega}{c}\,r^2\Big]\ , & \vphantom{\bigg|}\\
\text{with }\
	\tilde{a}=\frac12\bigg(1+\nu+\bfrac{ic\tilde{k}^2}{2\omega}\bigg)\ . &
\end{align}

From the expansion of this solution near~$r\!=\!0$, we again extract the source and the VEV for $\phi$, from which we then derive an expression similar to~\eqref{scalarLifcorr}. This results in the following boundary two-point correlator of the operator dual to our more general scalar:
\begin{equation}
\big\langle\mathcal{O}_{\phi}(k)\,\mathcal{O}^*_{\phi}(-k)\big\rangle=
2i\Big(-\bfrac{i\omega}{c}\Big)^{\!\nu}\; \frac{\Gamma[1-\nu]}{\Gamma[\nu]}\, \frac{\Gamma\Big[\frac{1}{2}\Big(1+\nu+\frac{ic}{2\omega}(k^2-h\omega)\Big)\Big]}{\Gamma\Big[\frac{1}{2}\Big(1-\nu+\frac{ic}{2\omega}(k^2-h\omega)\Big)\Big]}\ .
\label{moregenscalcorr}
\end{equation}
This expression matches the pure Klein-Gordon correlator~\eqref{scalarLifcorr} for $h\!=\!0$ and $c\!=\!1$. Let us however see how the analytic structure is changed by $h>0$, and what happens in the Schr\"odinger limit $c\!\to\!\infty$. 

The poles of the correlator are situated where the argument of the Gamma function in the numerator is a negative integer. In the complex $\omega$-plane, these are located at
\begin{equation}
\omega = -\frac{i\,k^2}{\tfrac{2}{c}\big(2n+1+\nu\big) -ih}\ , \quad \text{ with }n\in \mathbb{N}\,.
\end{equation}
We can see immediately that for~$h\!=\!0$ and $c\!=\!1$ we retrieve the same poles as in the Lifshitz Klein-Gordon case~\eqref{immpoles}. For~$c\!\gg\!1$, instead, the imaginary part of the poles for small $n$ is very small, and so they lie close to the positive real axis (since $h$ is taken to be positive).
Note that taking~$h>0$ leads to a first order time derivative in the equation of motion \eqref{geneq} which implies that time reversal invariance, here taken to be purely a sign flip of the time coordinate without any action on the fields, is violated. In the strict $c\rightarrow\infty$ limit the structure of the zeros and poles of the correlator changes in a non-trivial manner as we will show next.

We can assume that in the Schr\"odinger $c\!\to\!\infty$ limit, $\frac{ic\,\tilde{k}^2}{4\omega}\!\to\!\infty$, so that we can use Stirling's formula, valid for $x\!\to\!\infty$,
\begin{equation}
\Gamma(x)\simeq e^{(x-\frac12)\log x-x+\frac12\log 2\pi}\ ,
\end{equation}
which implies
\begin{equation}
\frac{\Gamma(x+a)}{\Gamma(x+b)}\simeq x^{a-b}\ .
\label{gammaid}
\end{equation}
We thus get,
\begin{equation}	\label{Schrocorr}
\langle O_\phi(k) O_\phi^*(-k)\rangle = 2i \frac{\Gamma[1-\nu]}{\Gamma[\nu]}\, \left(\frac{k^2-h\omega}{4}\right)^\nu\ ,
\end{equation}
which agrees with \cite{Balasubramanian:2008dm}. Note that for $c\!=\!\infty$ the equation \eqref{geneqz2} reduces to a Bessel equation, as for a relativistic scalar in AdS. Hence the correlator is similar to the one in the conformal case, where however the Lorentz invariant $q^2=k^2-\omega^2$ is replaced by the Lifshitz covariant combination $k^2-h\omega$. 
There is a branch cut which we choose to be placed at those points for which the argument of the fractional power $0<\nu<1$ is negative, that is for real and positive $\omega\!\geq\!\frac{k^2}{h}$. Furthermore, there are no poles, and in particular no imaginary poles as in the Lifshitz Klein-Gordon case. The enhancement of symmetry we have for this specific case (allowing for time-reversal violation and looking at the Schr\"odinger limit) fixes the dispersion relation to a precise form, \emph{i.e.}~$\omega=h^{-1}k^2$, leaving no room for imaginary poles.

Before concluding our discussion of complex scalar probes on Lifshitz spacetimes, we consider one final case: $c=\infty$ and $z\!=\!4$. We saw that for $c=\infty$ and $z=2$ we obtain the case of a Schr\"odinger scalar. In the same limit but with $z=4$ we do not see a symmetry enhancement but we do find an example where we have analytic control of the solution. Since this is rare for values of $z$ different from 2 we pause here to study this special case in some detail. 

Consider equation~\eqref{geneq} with $c\!\to\!\infty$ and $z\!=\!4$. We obtain, for the Fourier transformed field,
\begin{equation}	\label{eomSchroscalar}
r\partial_r(r\partial_r\phi) -\tilde{d}\,r\partial_r\phi -(m^2 +k^2r^2 -h\omega\,r^4)\phi =0 \ ,
\end{equation}
which is again an equation of the confluent hypergeometric kind. 
Imposing the usual in-falling boundary conditions, we obtain the following solution: 
\begin{align}
\quad \phi(r,\omega,k) = C\: e^{i\frac{1}{2}\sqrt{h\omega}\,r^2} r^{\frac{d+3}{2}+\nu}\; \sfU\Big[\,\hat{a},1+\nu;-i\sqrt{h\omega}\,r^2\Big]\ , & \vphantom{\bigg|}\\
\text{with }\
\hat{a}=\frac12\bigg(1+\nu+\bfrac{ik^2}{2\sqrt{h\omega}}\bigg)\ . &
\end{align}

Repeating the procedure that allowed us to derive the correlator~\eqref{scalarLifcorr}, we can write the two-point function for this $z\!=\!4$ scalar as:
\begin{equation}
\big\langle\mathcal{O}_{\phi}(k)\,\mathcal{O}^*_{\phi}(-k)\big\rangle=
2i\left(-i\sqrt{h\omega}\right)^{\!\nu}\; \frac{\Gamma[1-\nu]}{\Gamma[\nu]}\, \frac{\Gamma\Big[\frac{1}{2}\Big(1+\nu+\frac{ik^2}{2\sqrt{h\omega}}\Big)\Big]}{\Gamma\Big[\frac{1}{2}\Big(1-\nu+\frac{ik^2}{2\sqrt{h\omega}}\Big)\Big]}\ .
\label{z4scalcorr}
\end{equation}
This correlator exhibits a branch cut, as well as poles on the negative real axis located at
\begin{equation}	
\omega=-\bfrac{k^4}{4h(2n +1 +\nu)^2}\ , \quad \text{ with }n\in \mathbb{N}\,.
\end{equation}
It is not straightforward to come with an interpretation of the modes associated with these poles, besides the obvious fact that the dispersion relations respect the $z\!=\!4$ Schr\"odinger scaling, and that they violate time-reversal invariance, as expected.

\section{Lifshitz holography for a massless vector}\label{vectorLif}

We now discuss conserved currents in a Lifshitz theory, that holographically correspond to massless vectors in the bulk. This is a set up that has received surprisingly little attention (see \cite{Edalati:2013tma} for a rather different set up, at finite density). As we will see, the correlators that we will be able to compute analytically display very similar features to the case of the Klein-Gordon scalar.

We consider from the outset the most general bulk action for a massless vector in a fixed Lifshitz background,
\begin{equation}	\label{SvectorLif}
	S=-\frac{1}{4}\int\!d^{d+1}x \sqrt{-g\,}\; \gamma^{\mu\nu} \Big(\gamma^{\rho\sigma}-\bfrac{2\kappa}{\beta^2}\, B^{\rho}B^{\sigma}\Big)F_{\mu\rho}F_{\nu\sigma}\ ,
\end{equation}
where $B^{\mu}$ and $\gamma^{\mu\nu}$ are defined in~\eqref{Lifvector} and~\eqref{orthometric}, and $\kappa\ge 0$ is a parameter which generalizes the action for the massless vector (by putting $\kappa$ to one, we recover the Maxwell action for a free massless vector). 

Using bulk gauge invariance $\delta A_\mu\!=\!\partial_\mu \alpha$, we can fix the radial component to vanish:~$A_r\!=\!0$. Hence, the residual gauge transformations will only depend on the boundary coordinates $t,x_i$. The equations of motion obtained by varying \eqref{SvectorLif} with respect to $A_r$, $A_t$, and $A_i$ respectively are
\begin{align}
	&
	\kappa\, r^{2z}\,\partial_r\partial_tA_t -r^2\,\partial_r\partial_iA_i =0\ , \\
	&
	r^{d-z}\partial_r \big(r^{-d+z+2}\partial_r A_t\big)+r^{2}\big(\partial_i\partial_i A_t -\partial_t\partial_i A_i\big) =0\ , \\
	&
	r^{d+z-2}\partial_r \big(r^{-d-z+4}\partial_rA_i\big) -\kappa \, r^{2z}\big(\partial_t^2A_i-\partial_i\partial_tA_t\big) +r^{2}\big(\partial_j\partial_jA_i-\partial_i\partial_jA_j\big)=0\ .
\end{align}
We will split the vector $A_i$ into a transverse and a longitudinal part:\footnote{%
	Note that considering a transverse splitting over all boundary coordinates would be very inconvenient when $z\neq 1$, since $\partial_mA^m= -r^{2z}\partial_t A_t+r^2 \partial_i A_i$, i.e.~the $r$-dependence does not factorize.
}
\begin{equation}
	A_i=T_i+\partial_iL\ , \quad \text{with }\ \partial_i T_i=0\ . \label{vectorsplit}
\end{equation}
Under the residual gauge transformations, we have $\delta T_i=0$, $\delta L=\alpha$, and $\delta A_t=\partial_t\alpha$. The splitting leads to the following four equations:
\begin{align}
	&
	\kappa\, r^{2z}\,\partial_r\partial_tA_t -r^2\,\partial_r \partial_i^2L=0\ ,\label{constr} \\
	&
	r^{d-z}\partial_r\big(r^{-d+z+2}\partial_r A_t\big)+r^{2}\,\partial_i^2\big(A_t -\partial_t L\big)=0\ , \label{EomAt} \\
	&
	r^{d+z-2}\partial_r\big(r^{-d-z+4}\partial_r\partial_iL\big) +\kappa\,r^{2z}\,\partial_i\partial_t\big(A_t-\partial_tL\big) =0\ , \label{EomAL} \\
	&
	r^{d+z-2}\partial_r\big(r^{-d-z+4}\partial_rT_i\big) -\kappa\,r^{2z}\,\partial_t^2T_i +r^{2}\,\partial_j^2T_i=0 \label{EomAT}  \ .
\end{align}
Every term in the equations above is gauge invariant.

Starting from \eqref{SvectorLif}, we now compute the renormalized action. The first step is to reduce the bulk action to a boundary action, using the equations of motion:
\begin{align}
	S &= \left.\int\!d^{d+1}x\ r^{-d-z} \frac{1}{2}\Big[ \kappa\, r^{2+2z} F_{rt}F_{rt}- r^4 F_{ri}F_{ri} \Big]\right|_\mathrm{on\text{-}shell} \nonumber\\
	& = -\int_{r=\epsilon}\!d^{d}x\ \frac{1}{2}\Big[ \kappa\, r^{-d+z+2} A_t\partial_rA_t- r^{-d-z+4} A_i\partial_rA_i \Big]\ .
\end{align}
We now use the split \eqref{vectorsplit} and the constraint \eqref{constr} to obtain
\begin{equation}\label{regS}
	S_\mathrm{reg} = -\frac{1}{2} \int_{r=\epsilon}\!d^{d}x \Big[ \kappa\, r^{-d+z+2} \big(A_t-\partial_tL\big)\partial_rA_t- r^{-d-z+4} T_i\partial_rT_i \Big]\ .
\end{equation}
All terms in the expression above are manifestly gauge invariant.

The next step is to see if there are any divergent terms, and if so, to add the appropriate counterterms to cancel the divergences. This procedure depends on the specific values of $d$ and $z$. We will indicate for which values of $d$ and $z$ our counterterms are valid as we go along.

Let us first consider the timelike/longitudinal part. The purely radial solutions of the equations of motion~\eqref{constr}--\eqref{EomAL} are respectively $A_t\propto 1,\, r^{d-z-1}$ and~$L\propto 1,\, r^{d+z-3}$. Hence, from the equations of motion and assuming $z\geq 1$,  we expand $A_t$ and $L$ as
\begin{align}\label{Atexp}
	A_t&= a_0+a_1r^2+\;\ldots\; +\tilde{a}_0 r^{d-z-1}+\;\ldots\; \\
	L&= l_0+ l_1 r^{2z}+\;\ldots\; + \tilde{l}_0 r^{d+z-3} +\;\ldots\;   
\end{align}
where all coefficients depend on the boundary coordinates $t,x_i$ and where we included the first order corrections with the coefficients $a_1$ and $l_1$ as well. Under the residual gauge transformations we have that $\delta a_0=\partial_t\alpha$ and $\delta l_0=\alpha$. To avoid complications (\emph{i.e.} logarithmic terms and/or alternative quantization, as for instance in \cite{Argurio:2016xih}), we assume $d-z-1>0$ (equivalently $\tilde d>2z$), which since $z\geq 1$ also implies $d+z-3>0$. Let us notice that the constraint \eqref{constr} imposes
\begin{equation}
	\partial_t a_1= \frac{z}{\kappa}\, \partial_i^2 l_1\ , \qquad 
	\kappa\,(d-z-1) \partial_t \tilde{a}_0 = (d+z-3)\partial_i^2\tilde{l}_0\ ,
\end{equation}
and that \eqref{EomAt} and \eqref{EomAL} imply
\begin{equation}\label{subl}
a_1=\frac{\partial_i^2\left(a_0-\partial_tl_0\right)}{2(d-z-3)}\,,\qquad\partial_i l_1=\frac{\kappa}{z}\frac{\partial_t\partial_i\left(a_0-\partial_t l_0\right)}{2(d-z-3)}\,.
\end{equation}

The regularized action is
\begin{equation}
S_\mathrm{reg}^{t/L} = -\kappa\int_{r=\epsilon}\!\!d^{d}x\; \Big[
	r^{-d+z+3}\big(a_0-\partial_tl_0\big)\,a_1 +\;\ldots\; +\bfrac12(d-z-1)\big(a_0-\partial_tl_0\big)\, \tilde{a}_0\Big]\, ,
\end{equation}
where the superscript $t/L$ means that we ignore the second $T^i$ dependent term in \eqref{regS}. The first term is divergent if $d-z>\!3$. In that case, the dots represent other possibly (less) divergent terms. We note that if $d-z=\!3$, there would have been log-divergences that would have to be taken care of. The last term is finite and is the only one involving $\tilde{a}_0$ as all other terms with $\tilde{a}_0$ vanish as~$r\!\to\!0$.

If the first term diverges, we have to add a proper local, gauge invariant counterterm, which in this case is
\begin{equation}
	S^{t/L}_\mathrm{ct} = -\int_{r=\epsilon}\!d^{d}x \sqrt{-\hat{g}\,} \left[ \frac{\kappa}{2(d-z-3)}F_{ti}F^{ti} \right]\ , \label{ctcov}
\end{equation}
where $\hat{g}$ is the induced metric on the boundary. Using the expression for $a_1$ given in equation \eqref{subl}
we obtain the `renormalized' action for this sector:
\begin{equation}	\label{renActionTimeLong}
\begin{aligned}
S_\mathrm{ren}^{t/L}=S_\mathrm{reg}^{t/L}-S^{t/L}_\mathrm{ct}= 
	-\kappa\int_{r=\epsilon} d^{d}x\ \bigg[ \frac12 (d-z-1) \big(a_0-\partial_tl_0\big)\tilde{a}_0\: +\;\; & \\
		+\bfrac{r^{-d+z+3}}{2(d-z-3)}\, \partial_t T_i\partial_t T_i & \bigg]\ .
\end{aligned}
\end{equation}
We observe that our counterterm succeeds in suppressing the divergent terms in the timelike/longitudinal sector without adding finite terms, but that it leads to an additional term in the transverse sector. This term is divergent, but, as we will see in a moment, it will be cancelled by the counterterm in the transverse sector.

So, let us concentrate on the transverse sector. The purely radial solutions of the equation of motion~\eqref{EomAT} is given by $T_i\!\propto 1,\,r^{d+z-3}$. By using \eqref{EomAT}, we see that $T_i$ can be expanded near the boundary as
\begin{equation}
T_i = t_0 +r^2\,t_1 + \;\ldots\; +r^{2z}\,t_z + \;\ldots\; + r^{d+z-3}\,\tilde{t}_0 + \;\ldots \ ,
\end{equation}
where we are omitting the $i$-index on the coefficients and where we have
\begin{equation}\label{relation}
t_1= \frac{1}{2(d+z-5)}\partial_i^2 t_0\ ,\qquad t_z= -\bfrac{\kappa}{2z(d-z-3)}\partial_t^2 t_0\ .
\end{equation}
Again, as a matter of simplicity (no logarithmic terms), we will consider $d+z-3\neq0,\,2,\,2z$. The boundary action is then
\begin{equation}	\label{boundaryActionTransverse}	
S_\mathrm{on\text{-}shell}^T = \int_{r=\epsilon} d^{d}x\ \Big[  r^{-d-z+5}\, t_0 t_1+\;\ldots\;+z\,r^{-d+z+3}\, t_0 t_z +\;\ldots\;+\frac12 (d+z-3)\, t_0\tilde{t}_0 \Big]\ ,
\end{equation}
where the superscript $T$ means that we ignore the first time/longitudinal term in \eqref{regS}.

If the first term in \eqref{boundaryActionTransverse} diverges, it can be cancelled by the following counterterm
\begin{equation}	\label{ctnoncov}
	S^T_\mathrm{ct} = \int_{r=\epsilon}\!d^{d}x\ \sqrt{-\hat{g}\,}\: \bigg[ -\frac{1}{4}\frac{1}{d+z-5}\: F_{ij}F^{ij} \bigg]\ ,
\end{equation}
which does not generate any finite term containing $\tilde{t}_0$. Further, if the second term in \eqref{boundaryActionTransverse} diverges, this part will be cancelled by the second (transverse) term in \eqref{renActionTimeLong}. Therefore, by combining the counterterms of the two sectors, we obtain the fully renormalized action
\begin{align}
S_\mathrm{ren} &= S_\mathrm{on\text{-}shell}-S_\mathrm{ct}^{t/L}-S_\mathrm{ct}^{T} = \nn 
&=
	\int\!d^{d}x\ \bigg[\, 
		\frac12 (d+z-3)\,t_0\tilde{t}_0 -\frac12 \kappa(d-z-1)\big(a_0-\partial_tl_0\big)\,\tilde{a}_0 \bigg]\ .
\end{align}
Note that as soon as $z\neq1$  the two counterterms \eqref{ctcov} and \eqref{ctnoncov} are of the general form that we used in the action \eqref{SvectorLif}.

Solving the equations of motion for the fluctuations with some bulk boundary conditions would allow us to express~$\tilde{a}_0$ and~$\tilde{t}_0$ in terms of the gauge invariant combinations of the sources, $(a_0-\partial_tl_0)$ and $t_0$ respectively, through some non-local function. The sources for the currents are
\begin{equation}
	S_\mathrm{sources}= \int d^{d}x  \left\{ -a_0J_t -l_0\partial_i J_i +t_{i0}J_i^T\right\}\ .
\end{equation}
This results into
\begin{align}
\big\langle J_t J_t \big\rangle &= 
	i\,\kappa \left(d-z-1\right) \bfrac{\delta\tilde{a}_0}{\delta a_0}\ , \label{JtJtCorr} \\
\big\langle J_i^T J_j^T \big\rangle &= 
	-i\,(d+z-3)\bigg(\delta_{ij}-\bfrac{\partial_i\partial_j}{\partial_{k}^2}\bigg) \bfrac{\delta\tilde{t}_0}{\delta{t}_0}\ ,	 \label{JiJiCorr}
\end{align}
and the correlators involving $\partial_i J_i$ being proportional to \eqref{JtJtCorr}.  Note that due to invariance under the residual gauge transformations there is an operator identity:
\begin{equation}
\bfrac{\delta S_\mathrm{ren}}{\delta{l_0}} = \partial_t \bfrac{\delta S_\mathrm{ren}}{\delta a_0}\quad \Leftrightarrow \quad\partial_iJ_i = \partial_tJ_t\ ,
\end{equation}
i.e.~the current $(-J_t,J_i)$ is conserved.

\subsection{Two-point function for a Lifshitz vector}	\label{vectorLif2p}

Since we have performed the holographic renormalization of the most general bulk action for a massless vector field and derived the formul\ae~\eqref{JtJtCorr}, \eqref{JiJiCorr} for the two-point functions of the charge densities and currents, we may try, for specific cases, to get analytic expressions for these correlators. They turn out to be very similar to the scalar case.

Again, when $z=2$ the equations for the fluctuations can be solved analytically. The conditions from the previous subsection for avoiding logarithmic terms imply that $d$ has to be even. Let us start with the transverse part. The Fourier transformed version of equation~\eqref{EomAT} reads (we write $T_i\equiv T$)
\begin{equation}
r^2 \partial_r^2T -\left(d-2\right)r\partial_r T +r^{4}\,\tilde{\omega}^2 T -r^2 k_i^2 T =0\ .\label{eqatvtric}
\end{equation}
where $\tilde{\omega}^2\!=\!\kappa\omega^{2}$. Again we have an equation of the confluent hypergeometric form, whose in-falling solution is
\begin{equation}
T = C_T\; e^{\frac{i}{2}\tilde{\omega} r^2}\sfU\bigg[\bfrac{ik^2}{4\tilde{\omega}}-\bfrac{d-3}4,-\bfrac{d-3}2 ; -i\tilde{\omega} r^2\bigg]\ .
\end{equation}
Using the expansion of $\sfU[a,b;x]$ for small small~$x$~\eqref{Usmallx}, from~\eqref{JiJiCorr} we obtain the correlator
\begin{equation}
	\big\langle J_i^T J_j^T\big\rangle = 2i\left(\delta_{ij}-\bfrac{k_ik_j}{k^2}\right) (-i\tilde{\omega})^{\frac{d-1}2} \bfrac{\Gamma\big[-\!\frac{d-3}2\big]\Gamma\Big[\frac{ik^2}{4\tilde{\omega}}+\frac{d+1}4\Big]}{\Gamma\big[\frac{d-1}2\big]\Gamma\Big[\frac{ik^2}{4\tilde{\omega}}-\frac{d-3}4\Big]}\ ,
	\label{jijj}
\end{equation}
for $d$ even. The scaling dimension is the correct one: taking into account that $[\tilde{\omega}]\!=\!2$, the correlator has dimension $d-1$, which is $2[J_i]-(d+1)$ given that $[J_i]\!=\!d$. We notice that it is similar to the scalar one~\eqref{scalarLifcorr}, and it also has analogous imaginary poles, located at
\begin{equation}
	\sqrt{\kappa}\,\omega=- \frac{i\,k^2}{4n+d+1}\ ,\qquad n\in\mathbb{Z}\,.
\end{equation}

Let us also investigate the $k\to0$ and $\omega \to 0$ limits of the correlator \eqref{jijj}. For $k\to0$ and fixed $\omega$, all the $\Gamma$-functions approach constants and the propagator is just proportional to a transverse projector times $\omega^{(d-1)/2}$. For $\omega \to 0$ (so, $\tilde{\omega} \to 0$) at fixed $k$, we have to employ Stirling's formula (see \eqref{gammaid}), leading to
\begin{equation}
	\langle J_i^T J_j^T\rangle \simeq 2i\left( \delta_{ij}-\bfrac{k_ik_j}{k^2}\right) \left(\frac{k^2}{4}\right)^{\frac{d-1}2}\frac{\Gamma\left(-\frac{d-3}2\right)}{\Gamma\left(\frac{d-1}2\right)}\ ,
\end{equation}
which is analogous to the correlator in AdS, as it would be straightforwardly inferred from setting~$\tilde{\omega}\!=\!0$ in the equation of motion~\eqref{eqatvtric}. Up to a prefactor both limits are fixed by scaling considerations and the fact that the correlator is proportional to a projector.

We can also consider the temporal/longitudinal sector. Let us start by establishing an ordinary differential equation for the temporal part. If we call  
\begin{equation}
\dot{A}\equiv \partial_tA_t\ , \qquad \dot{L}\equiv \partial_i\partial_iL\ ,
\end{equation}
then equation \eqref{constr}, the time derivative of \eqref{EomAt} and the spatial divergence of \eqref{EomAL} become
\begin{align}
& \kappa\, r^{-d+z+2}\partial_r\dot{A}-r^{-d-z+4} \partial_r \dot{L}=0\ , \\
& \partial_r\big(r^{-d+z+2}\,\partial_r \dot{A}\big)+r^{-d+z+2}\,\big(\partial_i\partial_i \dot{A} -\partial_t^2 \dot{L}\big)=0\ , \label{secondeq}\\
& \partial_r \big(r^{-d-z+4}\,\partial_r\dot{L}\big) +\kappa\, r^{-d+z+2}\,\big(\partial_i\partial_i\dot{A}-\partial_t^2\dot{L}\big) =0\  ,
\end{align}
the third being obviously redundant. The first equation gives
\begin{equation}
\partial_r \dot{L}= \kappa\, r^{2z-2}\partial_r\dot{A}\ ,
\end{equation}
so that by multiplying \eqref{secondeq} with $r^{d-z-2}$ and differentiating with respect to $r$ we eventually have an equation for $\dot{A}$ only:
\begin{equation}
r^2\partial_r \Big( r^{d-z-2}\partial_r\big(r^{-d+z+2}\partial_r \dot{A}\big)\Big) - \kappa\, r^{2z}\partial_t^2\partial_r\dot{A}+r^2 \partial_i^2\partial_r\dot{A} = 0\ .
\end{equation}
This is a second order equation for $\dot{A}'\equiv\partial_r\dot{A}$:
\begin{equation}
r^2\partial_r^2 \dot{A}'-(d-z-2)r\partial_r \dot{A}'+(d-z-2)\dot{A}' -\kappa\, r^{2z}\partial_t^2 \dot{A}'+r^2 \partial_i\partial_i \dot{A}'=0\ . \label{Adot}
\end{equation}
Performing a Fourier transform and setting $z=2$, the latter equation becomes
\begin{equation}
r^2\partial_r^2 \dot{A}'-\left(d-4\right)r\partial_r \dot{A}'+\left(d-4\right)\dot{A}' +r^{4}\tilde{\omega}^2 \dot{A}'-r^2 k^2 \dot{A}'=0\ ,
\end{equation}
which is again of the confluent hypergeometric form. Thus, we obtain
\begin{equation}
\dot{A}'= C_T\; e^{\frac{i}{2}\tilde{\omega}\,r^2}\,r\; \sfU\bigg[\bfrac{ik^2}{4\tilde{\omega}}-\frac{d-7}4,-\frac{d-7}2;\,-i\tilde{\omega}\,r^2\bigg]\ .
\end{equation}
One must be careful with the expansions, because leading and subleading orders can get inverted. This happens for example for $d=4$ where the $\tilde{a}_0$ term is more leading than the $a_1$ term, with both being subleading to the $a_0$ term.

The $a_1$ and $\tilde a_0$ coefficients are given by:
\begin{equation}
2\omega\,a_1=iC_T\,\bfrac{\Gamma\big[\frac{d-5}2\big]}{\Gamma\Big[\bfrac{ik^2}{4\tilde{\omega}}+\frac{d-3}4\Big]}\ , \qquad 
(d-3)\omega\,\tilde{a}_0 = iC_T\,(-i\tilde{\omega})^{\frac{d-5}{2}}\bfrac{\Gamma\big[-\!\frac{d-5}2\big]}{\Gamma\Big[\bfrac{ik^2}{4\tilde{\omega}}-\frac{d-7}4\Big]}\,,
\end{equation}
so that, using \eqref{subl}, we get
\begin{equation}
\tilde{a}_0 =
	-\frac{1}{2(d-3)}\, k^2 (-i\tilde{\omega})^{\frac{d-3}{2}} \bfrac{\Gamma\big[\!-\!\frac{d-5}2\big]\Gamma\Big[\bfrac{ik^2}{4\tilde{\omega}}+\frac{d-3}4\Big]}{\Gamma\big[\frac{d-5}2\big]\Gamma\Big[\bfrac{ik^2}{4\tilde{\omega}}-\frac{d-7}4\Big]} \left(a_0+i\omega l_0\right)\ .
\end{equation}
Eventually, from~\eqref{JtJtCorr}, 
\begin{equation}
\big\langle J_t(k) J_t(-k) \big\rangle = 
	-i\,\frac{\kappa}{2}\; k^2 (-i\tilde{\omega})^{\frac{d-5}{2}}\; \bfrac{\Gamma\big[\!-\!\frac{d-5}2\big]\Gamma\Big[\bfrac{ik^2}{4\tilde{\omega}}+\frac{d-3}4\Big]}{\Gamma\big[\frac{d-3}2\big]\Gamma\Big[\bfrac{ik^2}{4\tilde{\omega}}-\frac{d-7}4\Big]}\ .
\label{jtjt}
\end{equation}
One expects the above correlator to have dimension $2[J_t]-(d+1)=d-3$, which it has. It also displays poles for
\begin{equation}
\sqrt{\kappa}\,\omega=- \frac{i\,k^2}{4n+d-3}\ .
\end{equation}

In the $\omega\to0$ limit, the correlator~\eqref{jtjt} becomes
\begin{equation}
\big\langle J_t(k) J_t(-k) \big\rangle = -2i\,\kappa\, \left(\frac{k^2}{4}\right)^{\!\frac{d-3}{2}}\bfrac{\Gamma\big[-\frac{d-5}2\big]}{\Gamma\big[\frac{d-3}2\big]}\ , 
\end{equation}
which is finite. It is to be noted that this is in line with the relativistic case, where the correlators of a conserved current are finite for $\omega\to 0$ and fixed $k$.

We have thus shown that a general massless vector in a Lifshitz background has two-point correlators that are very similar to the ones produced by a Klein-Gordon scalar in the same background. They display the same analytic structure, comprising a cut ending at the origin and an accumulation of poles on the lower imaginary axis.

\subsection{Galilean invariant probe action}
\label{GEDsection}
In the case of the most general action of the complex scalar \eqref{Smoregenscal} we noticed that there is a special case in which the probe action has Schr\"odinger symmetry. This in particular means that the probe action has Galilean boost invariance even though the Lifshitz background does not have a Killing vector generating Galilean boosts. The mechanism that makes this possible was uncovered in \cite{Hartong:2015wxa}. On the other hand the most general action for a massless vector \eqref{SvectorLif} does not have a special case for which the action enjoys Galilean boost invariance. In this subsection we will show that by adding a real scalar field we can write a probe action for a massless vector coupled to a real scalar that has the same symmetries as Galilean electrodynamics studied in \cite{Festuccia:2016caf}.

Consider the following probe action:
\begin{equation}
S=\int d^{d+1}x\sqrt{-g}\left(-\frac{1}{4}\gamma^{\mu\nu}\gamma^{\rho\sigma}F_{\mu\rho}F_{\nu\sigma}+\frac{1}{2c^2}B^\mu\partial_\mu\varphi B^\nu\partial_\nu\varphi-h\gamma^{\mu\nu}B^\rho F_{\mu\rho}\partial_\nu\varphi\right)\ ,
\end{equation}
where $\varphi$ is a real scalar. If we assume the background metric and massive vector field are of the usual form for a Lifshitz spacetime then we find
\begin{align}
S=\int d^{d+1}x\; r^{-z-d}\Big[&
	-\bfrac{r^4}{4}F_{ij}F_{ij}-\bfrac{r^4}{2}F_{ir}F_{ir}\, + \nn
&
	+\bfrac{\beta^2}{2c^2}r^{2z}\left(\partial_t\varphi\right)^2+h\beta r^{z+2}\big(F_{it}\partial_i\varphi+F_{rt}\partial_r\varphi\big) \Big]\ .
\end{align}
We now take $z=2$ so that $\beta=1$ and we furthermore take $c=1$ and $h=-1$, so that we obtain
\begin{equation}\label{GED}
S=\int d^{d+1}x\; r^{2-d}\left(-\frac{1}{4}F_{ij}F_{ij}-\frac{1}{2}F_{ir}F_{ir}+\frac{1}{2}\left(\partial_t\varphi\right)^2-F_{it}\partial_i\varphi-F_{rt}\partial_r\varphi\right)\ .
\end{equation}

We will now study the symmetries of \eqref{GED}. As advocated it has a Galilean symmetry. This symmetry is realized by the following transformations:
\begin{equation}
t'=t\,,\qquad x'^i=x^i+v^i t\,,\qquad r'=r\,,
\end{equation}
provided we transform the fields as
\begin{eqnarray}
A'_r & = & A_r\,,\\
A'_i & = & A_i+v^i\varphi\,,\\
A'_t & = & A_t-v^iA_i-\bfrac{1}{2}v^2\varphi\,,\\
\varphi' & = & \varphi\,.
\end{eqnarray}
We have thus found a Lifshitz counterpart of the GED action given in \cite{Festuccia:2016caf}. It furthermore has two independent scale symmetries. The first is given by
\begin{equation}
t\rightarrow\lambda t\,,\qquad x^i\rightarrow x^i\,,\qquad r'\rightarrow r\,,
\end{equation}
with the following field transformations:
\begin{equation}
A_r\rightarrow\lambda^{-1/2}A_r\,,\qquad A_i\rightarrow\lambda^{-1/2}A_i\,,\qquad A_t\rightarrow\lambda^{-3/2}A_t\,,\qquad\varphi\rightarrow\lambda^{1/2}\varphi\,.
\end{equation}
The second scale symmetry is given by
\begin{equation}
t\rightarrow t\,,\qquad x^i\rightarrow\mu x^i\,,\qquad r'\rightarrow\mu r\,,
\end{equation}
with the following field transformations:
\begin{equation}
A_r\rightarrow A_r\,,\qquad A_i\rightarrow A_i\,,\qquad A_t\rightarrow\mu A_t\,,\qquad\varphi\rightarrow\mu^{-1}\varphi\,.
\end{equation}
Another way of phrasing this result is that the GED probe action on a $z=2$ Lifshitz background has a Lifshitz scale symmetry for arbitrary values of a dynamical critical exponent which is not the $z$ of the background geometry. The only other symmetries of \eqref{GED} are time and space translations and spatial rotations.

Since it is necessary to introduce an additional real scalar in order to enhance the symmetry in this particular case, for simplicity in the remainder of this paper we will work with the vector-only model given in \eqref{SvectorLif}. In particular in the next section we will study the minimal coupling of \eqref{SvectorLif} to \eqref{Smoregenscal}.

\section{Lifshitz holography for a vector and a charged scalar}\label{GoldstoneLif}

We are finally ready to study the physics of symmetry breaking for a Lifshitz invariant theory in a holographic setup. Our main goal is to retrieve the correct non-relativistic Ward identities for symmetry breaking.

To this end we will consider the most general $\mathrm{U}(1)$-invariant action for a complex scalar field coupled to a massless vector field on a non-dynamical Lifshitz background. This amounts to combining the actions for the most general scalar~\eqref{Smoregenscal} and for the most general massless vector~\eqref{SvectorLif}, and replacing ordinary derivatives with covariant derivatives, leading to:
\begin{align}	\label{S-U1}
S=\int\!d^{d+1}x\; \sqrt{-g\,}\, \bigg[&
	-\frac{1}{4}\,\gamma^{\mu\nu} \Big(\gamma^{\rho\sigma}-\bfrac{2\kappa}{\beta^2}\,B^{\rho}B^{\sigma}\Big) F_{\mu\rho}F_{\nu\sigma} -\gamma^{\mu\nu}D_\mu\phi^*D_\nu\phi \;+  \\
&
	+\bfrac{1}{c^2\beta^2}\,B^{\mu}D_\mu\phi^*B^{\nu}D_\nu\phi -\bfrac{ih}{2\beta}\,B^\mu\Big(\phi^*D_\mu\phi-\phi{D_\mu}\phi^*\Big) -m^2\phi^*\phi \,\bigg] \ , \nonumber
\end{align}
where $\gamma_{\mu\nu}$ and~$B_\mu$ are defined in~\eqref{orthometric}, \eqref{Lifvector}, and 
\begin{align*}
F_{\mu\nu} 	& =	\partial_{\mu}A_{\nu}-\partial_{\nu}A_{\mu} \ ,\\
D_{\mu} 	& =	\partial_{\mu}-iA_{\mu} \ .
\end{align*}
Note that we have fixed the electric charge to unity, since we will not need to make it explicit.

We will partially fix the gauge freedom by imposing radial gauge, i.e. $A_r\!=\!0$, like we did in Section~\ref{vectorLif}. The equations of motion, obtained by varying the action~\eqref{S-U1} with respect to~$A_r$, $A_i$, $A_t$, and~$\phi^*$ respectively, are
\begin{align}
&	
	r^2\partial_i\partial_rA_i -\kappa\,r^{2z}\partial_t\partial_rA_t +i\,\big(\phi^*\partial_r\phi-\phi\partial_r\phi^*\big) =0 \ , \vphantom{\frac{}{|}} \label{EOMVecScaConstraint}\\
&
	r\partial_r\big(r\partial_rA_i\big) -\big(\tilde{d}-2\big)\,r\partial_rA_i +r^2\left(\partial_j^2A_i-\partial_i\partial_jA_j\right) -\kappa\,r^{2z}\left(\partial_t^2A_i-\partial_i\partial_tA_t\right) \, + \nn
&\qquad\qquad
		-i\left(\phi^*\partial_i\phi-\phi\partial_i\phi^*\right) -2\phi^*\phi A_i =0 \ , \vphantom{\frac{}{|}} \label{EOMVecScaAi} \\
&	
	r\partial_r\big(r\partial_rA_t\big) -\big(\tilde{d}-2z\big)\,r\partial_rA_t +r^2\left(\partial_i^2A_t-\partial_t\partial_iA_i\right) \, + \nn
&\qquad\qquad
		-\bfrac{i}{\kappa c^2}\left(\phi^*\partial_t\phi-\phi\partial_t\phi^*\right) -\bfrac{2}{\kappa c^2}\,\phi^*\phi A_t +\bfrac{h}{\kappa}\,r^{-z}\phi^*\phi =0 \ ,	\vphantom{\frac{}{|}} \label{EOMVecScaAt}\\
&
	r\partial_r\big(r\partial_r\phi\big) -\tilde{d}\,r\partial_r\phi -m^{2}\phi +r^2\left(\partial_i^2\phi-2iA_i\partial_i\phi-i(\partial_iA_i)\phi -A_i^2\phi\right)\, +  \nn
&\qquad\qquad
		-\bfrac{r^{2z}}{c^{2}}\left(\partial_t^2\phi-2iA_t\partial_t\phi-i(\partial_{t}A_{t})\phi-A_t^2\phi\right) -ih\,r^z\big(\partial_{t}\phi-iA_{t}\phi\big) =0 \label{EOMVecScaPhi} \ ,
\end{align}
where we remind the reader that $\tilde d=d+z-1$.

In order to trigger the breaking of the U(1) global symmetry on the boundary, we assign a background profile $\phi_B$ to the scalar, which can be taken to be real by virtue of the symmetry. In order not to break boundary spacetime symmetries, we will construct a set of $(t,x^{i})$-independent solutions $(\phi_B, A_{Bt}, A_{Bi})$ of the equations of motion where $A_{Bi}=0$ and $\phi_B \in \mathbb{R}$. As we will see a posteriori, we have to consider $A_{Bt}\neq 0$ otherwise the scalar background has to be zero.  With these specific requirements for the solutions, equations \eqref{EOMVecScaConstraint} and \eqref{EOMVecScaAi} are trivially satisfied and \eqref{EOMVecScaAt}, \eqref{EOMVecScaPhi} provide respectively:
\begin{align}
&	
	r\partial_r\big(r\partial_rA_{Bt}\big) -\big(\tilde{d}-2z\big)\,r\partial_rA_{Bt} 
	 -\bfrac{2}{\kappa c^2}\,\phi_{B}^{2} A_{Bt} +\bfrac{h}{\kappa}\,r^{-z}\phi_{B}^{2} =0 \ ,	\vphantom{\frac{}{|}} \label{BackEOMVecScaAt}\\
&
	r\partial_r\big(r\partial_r\phi_{B}\big) -\tilde{d}\,r\partial_r\phi_{B} -m^{2}\phi_{B} 
		+\bfrac{r^{2z}}{c^{2}}A_{Bt}^2\phi_{B} -h\,r^z A_{Bt}\phi_{B} =0 \label{BackEOMVecScaPhi} \ .
\end{align}
As we can see, whenever $h\neq 0$, the last term of \eqref{BackEOMVecScaAt} imposes $A_{Bt}\neq 0$ for~$\phi_B$ to be non-zero. If $h=0$ we can set $A_{Bt}=0$ in which case the solution for $\phi_B$ is given by 
\begin{equation}\label{phi_B}
\phi_B=w\,r^{\frac{\tilde{d}}{2}-\nu} +v\,r^{\frac{\tilde{d}}{2}+\nu}
\end{equation}
where $\tilde{d}$ and $\nu$ are the same as defined in Section~\ref{2pscalarLif}, eq.~\eqref{asymphi}. As usual, in order to avoid logarithms we stay away from the BF bound and the unitarity bound. Further for simplicity with the holographic renormalization we require \mbox{$\nu\in (0,1)$}. When $h\neq 0$ we will assume that $A_{Bt}$ is sufficiently subleading in order that the terms containing $A_{Bt}$ in \eqref{BackEOMVecScaPhi} are subleading, so that \eqref{phi_B} is still correct sufficiently close to the boundary. Equation \eqref{BackEOMVecScaAt} contains an inhomogeneous term which is the last term. This means that the leading order of $A_{Bt}$ must be the same as that of $r^{-z}\phi_{B}^{2}$. This implies that the third term in \eqref{BackEOMVecScaAt} is subleading. We thus obtain for the scalar and vector background profiles the following expansions:
\begin{eqnarray}
\phi_B & = & w\,r^{\frac{\tilde{d}}{2}-\nu} +v\,r^{\frac{\tilde{d}}{2}+\nu}+\ldots\,,\label{phiB}\\
A_{Bt} & = & A_{B}\,r^{\tilde{d}-z-2\nu}+B_Br^{\tilde d-z}+C_Br^{\tilde d-z+2\nu}+\dots\,,\label{ABt}
\end{eqnarray}
with 
\begin{equation}
A_{B}\propto h w^2\,,\qquad B_B\propto hwv\,,\qquad C_B\propto hv^2\,.
\end{equation}
The dots in the expressions for $\phi_B$ and $A_{Bt}$ denote terms that are higher order in $w$ and $v$.
This value of $A_{Bt}$ is subleading with respect to the component in the expansion which would represent a VEV for the associated charge (i.e.~the term analoguous to $\tilde a_0$ in \eqref{Atexp}) if $z>2\nu$, which we assume from now on.

We will now treat the fields as small fluctuations around these background profiles, expanding the action up to second order in the fluctuations. We will split the scalar fluctuations into a  real and an imaginary part, i.e.
\begin{equation}
\phi=\frac{\phi_B+\rho+i\pi}{\sqrt{2}}\ .
\end{equation}
For the moment we keep both the leading mode with coefficient $w$ and the subleading mode with coefficient $v$ of the scalar profile, corresponding respectively to explicit and spontaneous breaking in ordinary quantization, and vice versa in alternative quantization. We can switch off the appropriate explicit breaking term when studying the case of spontaneous symmetry breaking. The fluctuations around the time component of the background gauge field will again be denoted by $A_t$. We hope that this will not lead to any confusion. Finally, we split the spatial component of the gauge field into a transverse and a longitudinal part,
\begin{equation*}
A_{i}=T_{i}+\partial_{i}L \ , \quad \text{with }\ \partial_iT_i=0 \ . 	\label{AT+iL}
\end{equation*}

Since we are interested in an action that is second order in the fluctuations, we need to expand the equations of motion up to first order in the fluctuations. Upon using the leading order equations of motion for the background fields the first order fluctuations satisfy the following equations:
\begin{align}
&
r^2\partial_i^2\,r\partial_rL -\kappa\,r^{2z}\partial_t\,r\partial_rA_t -\phi_B\,r\partial_r\pi+\pi\,r\partial_r\phi_B =0 \phantom{\frac{}{|}} \label{eqAu}\\
&
r\partial_r(r\partial_rT_i) -(\tilde{d}-2)\,r\partial_rT_i -\kappa\,r^{2z}\partial_t^2T_i +r^2\partial_j^2T_i -\phi_B^2 T_i =0 \phantom{\frac{}{|}} \label{eqT}\\
&
r\partial_r(r\partial_rL) -(\tilde{d}-2)\,r\partial_rL -\phi_B^2\,L +\kappa\,r^{2z}\partial_t\big(A_t-\partial_tL\big) +\phi_B\pi =0 \phantom{\frac{}{|}} \label{eqL} \\
&
r\partial_r(r\partial_rA_t) -(\tilde{d}-2z)\,r\partial_rA_t -\kappa^{-1}c^{-2}\,\phi_B^2\,A_t -2\kappa^{-1}c^{-2}\,\rho\phi_{B}\,A_{Bt}\, + \nn
&\qquad\qquad
+r^2\partial_j^2\big(A_t-\partial_tL\big)
+\kappa^{-1}c^{-2}\,\phi_B\,\partial_t\pi +\kappa^{-1}h\,r^{-z}\phi_{B}\rho =0 	\phantom{\frac{}{|}} \label{eqAt}\\
&
r\partial_r(r\partial_r\rho) -\tilde{d}\,r\partial_r\rho -\left(m^{2}-r^2\partial_j^2+c^{-2}r^{2z}\partial_t^2-c^{-2}r^{2z}A_{Bt}^{2}+h\,r^{z}A_{Bt}\right)\rho \, + \nn
&\qquad\qquad\qquad\qquad\qquad\qquad\qquad
-h\,r^z\left(A_t \phi_{B}-\partial_t\pi\right) =0	\phantom{\frac{}{|}} \label{eqrho} \\
&
r\partial_r(r\partial_r\pi) -\tilde{d}\,r\partial_r\pi -\left(m^{2}-r^2\partial_j^2+c^{-2}r^{2z}\partial_t^2-c^{-2}r^{2z}A_{Bt}^{2}+h\,r^{z}A_{Bt}\right)\pi \,+ \nn
&\qquad\qquad\qquad\qquad
-h\,r^z\partial_t\rho -\phi_B\big(r^2\partial_j^2L-c^{-2}r^{2z}\partial_tA_t\big)+2c^{-2}r^{2z}A_{Bt}\partial_{t}\rho =0 \label{eqpi}
\end{align}
We see that the equation for $T_i$ decouples, whereas we get a system of coupled equations for $\rho$, $\pi$, $L$, and $A_t$. If $h=0$, in which case we take $A_{Bt}=0$, the real part $\rho$ decouples as well.

Using these equations of motion the part of the action that is quadratic in the fluctuations can be put on-shell and reduced to a boundary term:
\begin{align}
S_\mathrm{on\text{-}shell} = \frac{1}{2}\int_{r=\epsilon}\!d^{d}x\, r^{-\tilde{d}}\, & \Big[\;
r^2T_i\,r\partial_rT_i -r^2L\,r\partial_r\partial_j^2L -\kappa\,r^{2z}A_t\,r\partial_rA_t \; + 	\nn
& -2\,\kappa\,r^{2z}A_{t}r\partial_{r}A_{Bt} +2\rho\,r\partial_r\phi_B +\rho\,r\partial_r\rho +\pi\,r\partial_r\pi \;\Big] 
\label{SchOS}
\end{align}
where the term containing $A_{Bt}$ actually vanishes in the near-boundary expansion of the fields by virtue of the assumption $z>2\nu$.

We now have to study the divergent pieces in the regularized action that need to be renormalized. Since we have already done this for the scalar and the vector when they are decoupled, we need only focus on the effects on the procedure of the presence of the background profiles $\Phi_B$ and $A_{Bt}$.

\subsection{Holographic renormalization and Ward identities}

We start by considering the components of the gauge field. From equations~(\ref{eqT}--\ref{eqAt}) we can see that the generic backgrounds~(\ref{phiB}--\ref{ABt}) impact the asymptotic expansions by the following terms:
\begin{align}
&
T_i= \ldots+ t_{i(B)}r^{\tilde d-2\nu} +\ldots \ , \\
&
L= \ldots +l_B r^{\tilde d-2\nu} +\ldots \ , \\
&
A_t= \ldots +h a_B r^{\tilde d-z-2\nu} + a_B' r^{\tilde d-2\nu}+\ldots \ ,
\end{align}
where the first term appearing in the expansion for $A_t$ above is present only when $h\neq 0$ (it is due to the last term in \eqref{eqAt}).

It is straightforward to see that none of these background-dependent coefficients will survive in  \eqref{SchOS}, provided we stick to the (simplifying) assumptions $\nu<1$ and $z>2\nu$. Therefore, the renormalization of the vector sector goes through exactly in the same way as for the free vector discussed in the previous section. Namely, there will be counterterms in the transverse sector if $\tilde d>4$ (i.e.~$d+z-5>0$), and in the timelike/longitudinal one if $\tilde d>2+2z$ (i.e.~$d-z-3>0$). In any case, such counterterms \eqref{ctcov} and \eqref{ctnoncov} do not affect the finite part of the action.

To complete the renormalization of the action~\eqref{SchOS}, now we only need to remove the scalar divergences. 
Note that the expansions for $\rho$ and $\pi$ are the same as in \eqref{asymphi}, even in presence of the background.
The scalar divergences are then cured by the following standard counterterm, which includes the effect of the background profile (but removing the zeroth order term)
\begin{align}\label{Sctmass}
S^{\phi}_\mathrm{ct} &= 
\bigg(\frac{\dt}{2}-\nu\bigg)\int_{r=\epsilon}\!\!d^{d}x\,\sqrt{-\hat{g}\,}\;\left(\phi^*\phi-\frac{\phi_B^2}{2}\right) \\
&	
= \frac12\bigg(\frac{\dt}{2}-\nu\bigg)\int_{r=\epsilon}\!\!d^{d}x\,\sqrt{-\hat{g}\,}\; \Big(\rho^2 +2\phi_B\rho +\pi^2 \,\Big) \ . 	\nonumber
\end{align}
The final renormalized action is thus 
\begin{align}	\label{Srenord}
S_\mathrm{ren} &= 
\frac{1}{2}\int\!d^{d}x\: \Big[\,(\tilde{d}-2)\,t_{i0}\tilde{t}_{i0} -(\tilde{d}-2)\,l_0\partial_i^2\tilde{l}_0 -\kappa\, (\tilde{d}-2z)\,a_0\tilde{a}_0 \; + \\
&\qquad\qquad\qquad\qquad\qquad\qquad\qquad\qquad
+2\,\nu\, \big(\rho_0\tilde{\rho}_0 +2v\,\rho_0 +\pi_0\tilde{\pi}_0\big)\Big] \nn\nonumber
&=
\frac{1}{2}\int\!d^{d}x\: \Big[\, (\tilde{d}-2)t_{i0}\tilde{t}_{i0} -\kappa\, (\tilde{d}-2z)\big(a_0-\partial_tl_0\big)\tilde{a}_0 \; + \\
&\qquad\qquad\quad\;\;
+2\,\nu\, \Big(\rho_0\tilde{\rho}_0 +\big(\pi_0-wl_0\big)\left(\tilde{\pi}_0-vl_0\right)+2v\,\big(\rho_0+\pi_0l_0-{\textstyle\frac12}w\,l_0l_0\big)\Big)\Big]\ , \nonumber
\end{align}
where to obtain the second line we have used the constraint~\eqref{eqAu}, which reads
\begin{equation}
(\tilde{d}-2)\,\partial_i^2\tilde{l}_0 -\kappa\, (\tilde{d}-2z)\,\partial_t\tilde{a}_0 +2\,\nu\left(v\pi_0-w\tilde{\pi}_0\right)=0\ .
\end{equation}

From this renormalized action it is straightforward to recognize the Ward identities for symmetry breaking, it is sufficient to express the action only in term of the gauge invariant combinations of the sources.
Recall the non-trivial gauge transformations are
\begin{equation}
\delta l_0 = \alpha\ ,\qquad \delta a_0 = \partial_t \alpha\ , \qquad
\delta \pi_0 = w\alpha \ , \qquad \delta \tilde \pi_0 = v \alpha\ .
\end{equation}
Through the following identifications:
\begin{equation}\label{nonlocalord}
\begin{aligned}
\tilde{t}_{i0} &= f_t(\Box)\,t_{i0}\ ,	\\
	\tilde{a}_0 &= f_a(\Box)\big(a_0-\partial_tl_0\big) +g_a(\Box)\big(\pi_0-wl_0\big) +h\,h_a(\Box)\rho_0\ , \\
	\tilde{\pi}_0-vl_0 &= f_\pi(\Box)\big(\pi_0-wl_0\big)+g_\pi(\Box)\big(a_0-\partial_tl_0\big) +h\,h_\pi(\Box)\rho_0\ ,	\\
		\tilde{\rho}_0 &= f_\rho(\Box)\,\rho_0 +h\,g_\rho(\Box)\big(a_0-\partial_tl_0\big) +h\,h_\rho(\Box)\big(\pi_0-wl_0\big)  \ ,	
\end{aligned}
\end{equation}
we obtain indeed
\begin{align}	\label{SrenordWI}
\!S_\mathrm{ren}=\frac{1}{2}\int\!d^{d}x \Big[&\:
	(\tilde{d}-2)t_{i0}f_t(\Box)\,t_{i0} -\kappa\, (\tilde{d}-2z)\big(a_0-\partial_tl_0\big)f_a(\Box)\big(a_0-\partial_tl_0\big) \: + \nn
&
	+\big(\pi_0-wl_0\big)\Big( 2\nu\,g_\pi(\Box) -\kappa\, (\tilde{d}-2z)\,g_a(\Box)\Big)\big(a_0-\partial_tl_0\big) \: + \nn
&
	+h\rho_0\,\Big( 2\nu\,g_\rho(\Box) -\kappa\,(\tilde{d}-2z)\,h_a(\Box)\Big)\big(a_0-\partial_tl_0\big) \: + \nn
&
	+2\nu\,\Big(\rho_0f_\rho(\Box)\,\rho_0 +2v\,\big(\rho_0+\pi_0l_0-{\textstyle\frac12}w\,l_0l_0\big)\Big) \: \nn
&
	+2h\nu\,\rho_0\Big(h_\pi(\Box)+h_\rho(\Box)\Big)\big(\pi_0-wl_0\big) \: + \nn
&
	+2\nu\,\big(\pi_0-wl_0\big)f_\pi(\Box)\big(\pi_0-wl_0\big)\Big) \Big]\ . 
\end{align}
We point out that the presence of~$h\neq0$, which couples~$\rho$ to~$\pi,A_t,L$, has forced us to introduce four additional non-local functions that are not present for~$h=0$ (see~\cite{Argurio:2015via} for a situation similar to this latter case). However, since there is no explicit dependence on~$h$ in the on-shell action~\eqref{SchOS}, the terms that are bilinear in~$\pi_0$ and $(a_0-\partial_tl_0)$ are the same both for~$h\neq0$ and~$h=0$. So, the Ward identities for symmetry breaking are smoothly recovered in both cases. Indeed, we have for instance
\begin{align}
\!\!\big\langle \partial_i J_i(x)\, \ImO(0) \big\rangle &= 
	i\frac{\delta^2 S_\mathrm{ren}}{\delta{l_0}\delta\pi_0} \\
&=
	i\Big[2\, \nu\big( v -w\,f_\pi(\Box)\big) +\Big(\nu\,g_\pi(\Box) -\bfrac{\kappa}{2}(\tilde{d}-2z)\,g_a(\Box)\Big)\partial_t\Big]\delta(x)\ , \nonumber\\
\big\langle J_t(x)\,\ImO(0) \big\rangle &= 
	i\frac{\delta^2 S_\mathrm{ren}}{\delta{a_0}\delta\pi_0} = i\,\Big(\nu\,g_\pi(\Box)-\bfrac{\kappa}{2}(\tilde{d}-2z)\,g_a(\Box)\Big)\,\delta(x)\  ; \\
\big\langle \ImO(x)\,\ImO(0) \big\rangle &= 
	-i\frac{\delta^2 S_\mathrm{ren}}{\delta{\pi_0}\delta\pi_0} = -i\,2\nu\,f_\pi(\Box)\,\delta(x)\  ;
\end{align}
and consequently
\begin{equation}
-\big\langle \partial_t J_t(x)\,\ImO(0) \big\rangle +\big\langle \partial_i J_i(x)\,\ImO(0) \big\rangle = i\,2\,\nu\,v \,\delta(x)\ +w\,\big\langle \ImO(x)\,\ImO(0) \big\rangle\ ,
\end{equation}
which is the usual Ward identity for concomitant spontaneous and explicit symmetry breaking. The other relations among correlators can be straightforwardly derived, and they follow from gauge invariance of $S_\mathrm{ren}$. Such relations were already derived with similar holographic techniques but in a relativistic context in~\cite{Argurio:2015wgr}.

We have thus shown how symmetry breaking in holographic Lifshitz field theories can be displayed at the level of the renormalized action. It would be interesting to push this effort further and solve, possibly in some specific model, for the correlators and then find the poles related to the (pseudo-)Goldstone modes, for instance in the Fourier transform of  $f_\pi(\Box)$. Unfortunately this involves solving a system of three (or even four if $h\neq0$) second order differential equations, plus a first order constraint. We postpone to future work the analytic and numerical study of this system.

\section{Conclusion}

The purpose of this paper was to establish the rules to discuss symmetry breaking in holographic theories with Lifshitz scaling. We have indeed written a renormalized action, that is the generating functional for the one- and two-point functions, which encodes the correct Ward identities for a field theory corresponding to a bulk action coupling a massless gauge vector to a charged complex scalar. We have allowed for the most general bulk action with such field content, that respects Lifshitz symmetry.\footnote{
	One can of course imagine additional self-interaction terms for the scalar, but those would not affect our quadratic renormalized action.}

Though we leave for further work a detailed study, through the correlators, of the physical spectrum of this model, it is already possible to infer the presence of Goldstone, i.e.~gapless, modes from the Ward identities in the purely spontaneous case, in the usual fashion (see the Appendix).

Along the way, we have also considered the cases of a free scalar and a free vector in the bulk. In this case, for specific values of $z$, it was possible to obtain analytic correlators, more specifically two-point functions of complex scalar operators and of conserved currents in the boundary field theory. We have thus shown that the known feature of scalar two-point functions, namely the accumulation of diffusive poles towards the origin of the $\omega$ plane, along the negative imaginary axis, persists both in more general scalar models and in the case of the current. It thus seems to be a generic feature of holographic Lifshitz theories.

Some simplifications occur when the symmetry is enhanced to a Schr\"odinger or Galilean one. In particular, the analytic correlator no longer displays the diffusive poles. It would be interesting to push further the holographic GED model discussed in Section \ref{GEDsection}, considering also the coupling to a charged scalar in the bulk. One would need to find the boundary dual interpretation of the real scalar field $\varphi$.

It would be nice to establish low-energy effective field theories which could reproduce, at least qualitatively, the correlators that we have derived in holography, and possibly investigate a situation of spontaneously broken symmetry. It should also be possible to discuss pseudo-Goldstone bosons and their masses for small explicit breaking, along the lines of~\cite{Argurio:2015wgr}.

\section*{Acknowledgements}

We would like to thank T.~Andrade, M.~Bertolini, N.~Iqbal, C.~Keeler, Y.~Korovin, J.~Tarrio and M.~Taylor for discussions.
This research has been supported in part by IISN-Belgium (convention 4.4503.15). R.A.~is
a Senior Research Associate of the Fonds de la Recherche Scientifique-F.N.R.S. (Belgium). During the initial phase the work
of J.H. was supported by a Marina Solvay fellowship as well as by the advanced ERC grant `Symmetries and Dualities in Gravity and M-theory' of Marc
Henneaux. J.H. thanks the ULB for its financial support and hospitality during the concluding phase of the research.

\appendix

\section{Ward identities in Lifshitz invariant field theories}
In this appendix we collect some results concerning Ward identities and Goldstone bosons in Lifshitz field theories. We start by deriving mixed correlators between currents and order parameters in low-energy effective field theories of Goldstone bosons. We then discuss how the qualitative features of these correlators can be extracted from the Ward identities. We finally comment on the relation between Lifshitz scaling and the presence of chemical potential in an otherwise relativistic field theory.

\subsection{Low energy theories for Goldstone bosons} 	\label{lowenGB}
Consider the low-energy effective action for a Goldstone boson in a field theory which enjoys Lifshitz scaling $t\to \lambda^z t$, $x_i\to \lambda x_i$, and which is invariant under time reflections~\cite{Griffin:2013dfa,Griffin:2014bta,Griffin:2015hxa,Horava:2016vkl}:
\begin{equation}
	S=\int dt d^{d-1}x\  \frac12\left( \partial_t \phi \partial_t \phi -(-1)^z\xi \phi \nabla^{2z} \phi \right) \ ,\label{action}
\end{equation}
where $\nabla^2=\partial_i\partial_i$, and the sign in front of the second term is chosen such that the dispersion relation reads $\omega^2=\xi k^{2z}$, so that we can set $\xi\geq 0$.

The relativistic case is $z=1$, $\xi=1$. It can be more reassuring to think of $z$ as an integer, but it can really take any value (here we will mainly consider $z\geq 1$). Assuming $\xi$ does not scale, the scaling dimensions are the following:
\begin{equation}
	[\partial_t]=z\ , \quad [\partial_i]=1 \ , \quad [\phi] = \frac{d-1-z}{2}\ .
\end{equation}
Note that for $d\leq z+1$ the scalar field has vanishing or negative scale dimension, which makes its fluctuations long range, rendering the effective action ill-defined. Note that this is equivalent to the holographic equally problematic case $\tilde d\leq 2z$, when the temporal component of the vector is at the BF bound, or needs alternative quantization (see \eqref{Atexp} and the discussion below).

The propagator for $\phi$ that one can extract from \eqref{action} is the following, in Fourier space:
\begin{equation}
	\langle \phi(\omega, q)\phi(-\omega,-q)\rangle = \frac{i}{\omega^2-\xi k^{2z}}\ . \label{prop}
\end{equation}
It can be checked that it has the correct scaling dimension.

The action \eqref{action} has a shift symmetry $\phi\to\phi+v\alpha$, with $v$ the VEV and $\alpha$ the parameter of the transformation. This is indeed expected for a Goldstone boson.

In order to find the current that generates this symmetry (which is broken by the VEV $v$) we promote $\alpha$ to a spacetime dependent function, and define
\begin{equation}
	\delta S = \int dt d^{d-1}x \left(\partial_t \alpha J_t - \partial_i \alpha J_i\right) \ .
\end{equation}
We then obtain
\begin{equation}
	J_t = v\partial_t \phi\ , \qquad J^i = (-1)^{z-1}\xi v \partial_i \nabla^{2z-2}\phi\ .
\end{equation}
They are linear, as it  befits currents of a broken symmetry (at the lowest order). 
The conservation law is
\begin{equation}
	\partial_t J_t -\partial_i J_i = v (\partial_t^2+(-1)^z\xi \nabla^{2z})\phi=0
\end{equation}
using the EOM. Note that it reads exactly as in the relativistic case, however the dimensions of the currents are now different:
\begin{equation}
	[J_t]=d-1\ , \qquad [J_i]= d+z-2\ .
\end{equation}
We can now check how the conservation law appears in two point functions, i.e. in the Ward identities. Recall that here the operator breaking the symmetry is $\phi$ itself, with $\langle\delta_\alpha \phi\rangle =v$.  

Using \eqref{prop}, we have
\begin{align}
	\langle J_t \phi \rangle &= -iv \omega \langle \phi \phi\rangle = \frac{v \omega}{\omega^2-\xi k^{2z}}\ ,\label{jt} \\
	\langle J_i \phi \rangle &= i\xi v k_i k^{2z-2}\langle \phi \phi\rangle = -\frac{\xi v k_i k^{2z-2}}{\omega^2-\xi k^{2z}}\ ,\label{ji}
\end{align}
so that 
\begin{equation}
	i\omega \langle J_t \phi \rangle + ik_i \langle J_i \phi \rangle = 
	\frac{iv \omega^2}{\omega^2-\xi k^{2z}}-\frac{i\xi v  k^{2z}}{\omega^2-\xi k^{2z}}=iv\ .
\end{equation}
This is the Ward identity
\begin{equation}
	-\partial_t\langle J_t \phi \rangle +\partial_i \langle J_i \phi \rangle =i\langle\delta_\alpha \phi\rangle\ .
\end{equation}

\subsubsection*{Forsaking T-invariance}

We can briefly consider the case of a theory which has Lifshitz scaling but not time reversal symmetry. The low energy action is then\footnote{%
	We need to write an action for a complex scalar, leading however to only one massless physical degree of freedom, the Goldstone boson, because of linearity in time derivatives. Note that we assume boundary terms that make the action real.}
\begin{equation}
	S=\int dt d^{d-1}x\  \left(  i\phi^* \partial_t \phi -(-i)^z\zeta \phi^* \nabla^{z} \phi \right) \ .\label{actionnont}
\end{equation}
Again, it is easier to consider $z$ even, but it can be more general. Now the dimension of the Goldstone field is 
\begin{equation}
	[\phi]= \frac{d-1}{2}\ .
\end{equation}
Note that it is always positive as long as $d>1$. 
Its propagator is 
\begin{equation}
	\langle \phi \phi^*\rangle =\frac{i}{\omega-\zeta k^z}\ .
\end{equation}
The currents read
\begin{equation}
	J_t = -iv \phi\ , \qquad J_i = -(-i)^z\zeta v \partial_i \nabla^{z-2}\phi\ .
\end{equation}
Their correlators are
\begin{align}
	\langle J_t \phi^* \rangle &=  \frac{v}{\omega-\zeta k^{z}}\ ,\label{jtnont} \\
	\langle J_i \phi^* \rangle & = -\frac{\zeta v k_i k^{z-2}}{\omega-\zeta k^{z}}\ ,\label{jinont}
\end{align}
so that the Ward identities are realized again
\begin{equation}
	i\omega \langle J_t \phi^* \rangle + ik_i \langle J_i \phi^* \rangle = 
	\frac{iv \omega}{\omega-\zeta k^{z}}-\frac{i\zeta v  k^{z}}{\omega-\zeta k^{z}}=iv\ .
\end{equation}

\subsection{From Ward identities to the Goldstone boson}
Having seen how the Ward identities are realized in the prototypical example of the low energy effective theory of the Goldstone bosons, 
we now reverse the logic and start from the Ward identities in order to find the Goldstone boson, i.e.~a low energy mode with gapless dispersion relation. We have
\begin{equation}
	-\partial_t\langle J_t \CO \rangle +\partial_i \langle J_i \CO \rangle =i\langle\CO\rangle\ ,
\end{equation}
for some operator which transforms under the symmetry generated by the currents, and which has a VEV that breaks the symmetry.

Using rotational symmetry, we parametrize the correlators in Fourier space as follows:
\begin{equation}
	\langle J_t \CO \rangle= f(\omega, k)\ , \qquad \langle J_i \CO \rangle = k_i g(\omega, k)\ .
\end{equation}
Note that $[f]=\Delta -z$ and $[g]= \Delta - 2$, where $\Delta$ is the dimension of the operator $\CO$. The Ward identity then implies
\begin{equation}
	\omega f +k^2 g = \langle\CO\rangle\ .
\end{equation}

Obviously, assuming $\langle\CO\rangle$ finite and non zero, when $\omega, k\to 0$, either $f$ or $g$, or both, have to blow up, signaling the presence of a massless particle in the spectrum, the Goldstone boson. 

Let us be more precise. Take first $k\to0$ with $\omega\neq 0$. Then, assuming $g$ finite in this limit, we have $f\to\frac{\langle\CO\rangle}{\omega}$. similarly, when $\omega\to 0$ at $k\neq0$, we have $g\to\frac{\langle\CO\rangle}{k^2}$. We can then rewrite
\begin{equation}
	f=\frac{\langle\CO\rangle}{\omega}\tilde f\ , 
\end{equation}
where $\tilde f$ is a dimensionless function of $\omega$ and $k$, and by virtue of the Ward identity
\begin{equation}
	g=\frac{\langle\CO\rangle}{k^2}(1-\tilde f)\ .
\end{equation}
There are two trivial ways to satisfy the Ward identity, which is setting either $\tilde f=1$ or $\tilde f=0$. 
These two choices do not corresponding to propagating degrees of freedom in the usual sense (i.e.~they lead to degenerate dispersion relations $\omega=0$ or $k^2=0$ respectively).
We thus consider the only interesting case where $\tilde f$ is a non-trivial function. Requiring that the low energy theory has Lifshitz scaling, then it must be a function of the ratio $x=\frac{k^z}{\omega}$. If we also impose time reversal symmetry, then it must be a function of $x^2$. The conditions on the $k\to 0$ and $\omega \to 0$ limits translate into
\begin{equation}
	\tilde f (x=0)=1\ , \qquad \tilde f(x=\infty)=0\ .
\end{equation}
We can readily find simple functions that satisfy the above requirements and reproduce the correlators obtained previously.
Without imposing time reversal symmetry, we can take 
\begin{equation}
	\tilde f =\frac{1}{1-\zeta x}
\end{equation}
so that
\begin{align}
	\langle J_t \CO \rangle&= \frac{\langle\CO\rangle}{\omega}\frac{1}{1-\zeta\frac{k^z}{\omega} }=\frac{\langle\CO\rangle}{\omega-\zeta k^z}\ , \\
	\langle J_i \CO \rangle &= k_i\frac{\langle\CO\rangle}{k^2}\left( 1- \frac{1}{1-\zeta\frac{q^z}{\omega} }\right) =-\frac{\zeta k_i k^{z-2} \langle\CO\rangle}{\omega-\zeta k^z} \ .
\end{align}
These have the same form as \eqref{jtnont}--\eqref{jinont}.

Imposing now time reversal invariance, we can take
\begin{equation}
	\tilde f =\frac{1}{1-\xi x^2}
\end{equation}
so that
\begin{align}
	\langle J_t \CO \rangle&= \frac{\langle\CO\rangle}{\omega}\frac{1}{1-\xi\frac{k^{2z}}{\omega^2} }=\frac{\omega\langle\CO\rangle}{\omega^2-\xi k^{2z}}\ , \\
	\langle J_i \CO \rangle &= k_i\frac{\langle\CO\rangle}{k^2}\left( 1- \frac{1}{1-\xi\frac{k^{2z}}{\omega^2} }\right) =-\frac{\xi k_i k^{2z-2} \langle\CO\rangle}{\omega^2-\xi k^{2z}} \ .
\end{align}
These have the same form as \eqref{jt}--\eqref{ji}.

Note that in both cases, more complicated functions can be taken. However, as soon as there is a denominator with a polynomial in $x$ (or $x^2$), near its roots the function will be very close to the ones we have taken. It would be nice to understand better from general principles what possible analytic structures $\tilde f$ can have.

\subsection{Type B Goldstone bosons as Lifshitz Goldstone bosons}

Here we aim at reviewing how Lifshitz-scaling Goldstone boson low-energy theories can emerge from a relativistic theory, when Lorentz boosts are broken. We consider a relativistic theory with a non-vanishing chemical potential
\cite{Miransky:2001tw,Schafer:2001bq}
\begin{equation}
	S=\int dt d^{d-1}x\left[ (\partial_t+i\mu) \phi (\partial_t-i\mu) \phi^* -\partial_i \phi \partial_i \phi^*-V(\phi) \right] \ .\label{chemS}
\end{equation}
Assuming $V(\phi)$ is such that there is symmetry breaking, we then split the real and imaginary parts and eventually obtain at quadratic order
\begin{equation}
	S=\frac12\int dt d^{d-1}x\left[ (\partial_t\phi_R)^2+ (\partial_t\phi_I)^2+2\mu 
	(\phi_R\partial_t\phi_I-\phi_I\partial_t\phi_R) -(\partial_i \phi_R)^2- (\partial_i \phi_I)^2-m^2 \phi_R^2 \right] \ .\label{chemSri}
\end{equation}
We assume that $V(\phi)$ is such that at most the real part gets a mass. For $m^2\neq 0$ this is a system similar to the one of type A Goldstone bosons, while for $m^2=0$ this is similar to type B Goldstone bosons (see for instance \cite{Watanabe:2012hr,Kapustin:2012cr,Argurio:2015via}).

Going to Fourier space, we see that the action can be rewritten as 
\begin{equation}
	S=\frac12\int d\omega d^{d-1}k\ (\begin{array}{cc} \phi_R & \phi_I
	\end{array}) \left( \begin{array}{cc} \omega^2 -k^2 -m^2 & 2i\mu\omega \\
		-2i\mu\omega & \omega^2 -k^2 
	\end{array}\right) \left(\begin{array}{c} \phi_R \\ \phi_I
	\end{array}\right)
\end{equation}
The dispersion relations can be read off from the zeros of the determinant of the above matrix:
\begin{equation}
	(\omega^2 -k^2 -m^2)(\omega^2 -k^2)-4\mu^2 \omega^2=0\ . \label{deteq}
\end{equation}
We always have a massive mode and a massless mode. The massive mode can be found setting $k=0$ above and has
\begin{equation}
	\omega^2 \simeq m^2+4\mu^2\ .
\end{equation}
Note that it is massive even when $m=0$. It will eventually not be part of the low-energy effective theory.  
The massless mode on the other hand depends on whether $m\neq0$ or not.
When $m\neq0$ \eqref{deteq} is approximated at low energies and momenta by $-m^2(\omega^2 -k^2)-4\mu^2 \omega^2=0$ and so
\begin{equation}
	\omega^2 \simeq \frac{m^2}{m^2+4\mu^2}k^2\ ,
\end{equation}
a gapless mode with a speed of propagation smaller than one, and which depends on $m$ and $\mu$. 
When $m=0$ \eqref{deteq} at low energies and momenta becomes $k^4-4\mu^2 \omega^2=0$, so that we have
\begin{equation}
	\omega^2 \simeq \frac{1}{4\mu^2}k^4\ ,
\end{equation}
a quadratic dispersion relation.

When $\mu$ is large with respect to the other scales in the problem, it makes sense to suppress the terms with two time derivatives. The effective action is 
\begin{equation}
	S_\mathrm{eff}=\int dt d^{d-1}x\left[ 2\mu 
	(\phi_R\partial_t\phi_I-\phi_I\partial_t\phi_R) -(\partial_i \phi_R)^2- (\partial_i \phi_I)^2-m^2 \phi_R^2 \right] \ ,\label{Seff}
\end{equation}
and the condition for a vanishing determinant becomes
\begin{equation}
	(k^2 +m^2)k^2-4\mu^2 \omega^2=0\ . \label{deteqeff}
\end{equation}
We see that we loose half of the modes, the massive ones as can be seen by setting $k=0$ above. We have effectively integrated them out. When $m\neq 0$, the massless mode has linear dispersion relation
\begin{equation}
	\omega^2 \simeq \frac{m^2}{4\mu^2}q^2\ ,
\end{equation}
which is the $\mu\gg m$ limit of the previous one. When $m=0$ the relation is unchanged. 

The latter case is can be reformulated in the following way. When $m=0$ there is no other scale in the problem and $\mu$ can be reabsorbed in the other variables. The effective action can be rewritten as 
\begin{equation}
	S_\mathrm{eff}=\int dt d^{d-1}x\left[ 
	\phi_R\partial_t\phi_I-\phi_I\partial_t\phi_R -(\partial_i \phi_R)^2- (\partial_i \phi_I)^2\right] \ ,\label{Seffeff}
\end{equation}
where now 
\begin{equation}
	[\partial_t]=2\ , \quad [\partial_i]=1\ , \quad [\phi]= \frac{d-1}{2}\ ,
\end{equation}
i.e.~this is precisely the $z=2$ Lifshitz theory in \eqref{actionnont}. It is now obvious that the dispersion relation of the type B Goldstone boson will be $\omega=k^2$, respecting $z=2$ Lifshitz scaling.

As a last comment, note that one can now turn on a mass operator, which is always relevant also in Lifshitz terms, but breaks the scaling symmetry. One then finds that the dispersion relation becomes $\omega=mk$. So even type A Goldstone bosons can be treated in the framework of (broken) Lifshitz theories.

It remains an open problem to obtain $z\neq 2$ Lifshitz scaling theories (such as the ones discussed in \cite{Watanabe:2014zza}, for instance) in the low-energy regime of relativistic ones.


\bibliography{lifshitzrefs}
\bibliographystyle{utphysmio}

\end{document}